\documentclass{article}

\usepackage{microtype}
\usepackage{graphicx}
\usepackage{subfigure}
\usepackage{booktabs} 

\usepackage{hyperref}

\usepackage[accepted]{icml2025}

\usepackage{amsmath}
\usepackage{amssymb}
\usepackage{mathtools}
\usepackage{amsthm}

\usepackage[capitalize,noabbrev]{cleveref}

\usepackage{multirow}
\usepackage{tabularx}

\theoremstyle{plain}
\newtheorem{theorem}{Theorem}[section]
\newtheorem{proposition}[theorem]{Proposition}
\newtheorem{lemma}[theorem]{Lemma}
\newtheorem{corollary}[theorem]{Corollary}
\theoremstyle{definition}
\newtheorem{definition}[theorem]{Definition}
\newtheorem{assumption}[theorem]{Assumption}
\theoremstyle{remark}

\newtheorem{example}{Example}
\newtheorem{heuristic}{Heuristic}

\usepackage[textsize=tiny]{todonotes}

\icmltitlerunning{Learning Mean Field Control on Sparse Graphs}

\begin{document}

\twocolumn[
\icmltitle{Learning Mean Field Control on Sparse Graphs}




\begin{icmlauthorlist}
\icmlauthor{Christian Fabian}{yyy,zzz}
\icmlauthor{Kai Cui}{yyy} 
\icmlauthor{Heinz Koeppl}{yyy,zzz}
\end{icmlauthorlist}

\icmlaffiliation{yyy}{Department of Electrical Engineering and Information Technology, Technische Universität Darmstadt, Germany}
\icmlaffiliation{zzz}{Hessian Center for Artificial Intelligence (hessian.AI)}

\icmlcorrespondingauthor{Heinz Koeppl}{heinz.koeppl@tu-darmstadt.de}

\icmlkeywords{Machine Learning, ICML}

\vskip 0.3in
]



\printAffiliationsAndNotice{}  

\begin{abstract}
    Large agent networks are abundant in applications and nature and pose difficult challenges in the field of multi-agent reinforcement learning (MARL) due to their computational and theoretical complexity. While graphon mean field games and their extensions provide efficient learning algorithms for dense and moderately sparse agent networks, the case of realistic sparser graphs remains largely unsolved. Thus, we propose a novel mean field control model inspired by local weak convergence to include sparse graphs such as power law networks with coefficients above two. Besides a theoretical analysis, we design scalable learning algorithms which apply to the challenging class of graph sequences with finite first moment. We compare our model and algorithms for various examples on synthetic and real world networks with mean field algorithms based on Lp graphons and graphexes. As it turns out, our approach outperforms existing methods in many examples and on various networks due to the special design aiming at an important, but so far hard to solve class of MARL problems.
\end{abstract}

\section{Introduction}

Despite the rapid developments in the field of multi-agent reinforcement learning (MARL) over the last years, systems with many agents remain hard to solve in general \citep{canese2021multi, gronauer2022multi}. Mean field games (MFGs) \citep{caines2006large, lasry2007mean} and mean field control (MFC) \cite{andersson2011maximum, bensoussan2013mean} are a promising way to model large agent problems in a computationally tractable manner and to provide a solid theoretical framework at the same time. While MFGs consider competitive agent populations, the focus of MFC are cooperative scenarios where agents optimize a common goal. The idea of MFC and MFGs is to abstract large, homogeneous crowds of small agents into a single probability distribution, the \emph{mean field} (MF). While MFC and MFGs have been used in various areas ranging from pedestrian flows \citep{bagagiolo2019optimal, Achdou2020} to finance \cite{carmona2018probabilistic, carmona2023deep} and oil production \citep{bauso2016robust}, the assumption of indistinguishable agents is not fulfilled in many applications.

A particularly important class of MARL problems are those with many connected agents. Initially, these agent networks were modeled by combining the graph theoretical concept of graphons \citep{lovasz2012large} with MFGs, resulting in graphon MFGs (GMFGs) \citep{caines2019graphon, caines2021graphon, cuilearning, zhang2024learning} and graphon MFC \cite{hu2023graphon}. Since GMFGs only model often unrealistic dense graphs, subsequently mean field models based on Lp graphons \citep{borgs2018lp, borgs2019L} and graphexes \citep{veitch2015class, caron2017sparse, borgs2018sparse} were developed, called LPGMFGs and GXMFGs, respectively \citep{fabian2023learning, fabian2024learning}.
While these models facilitate learning algorithms in moderately sparse networks, they exclude sparser topologies. Formally, (LP)GMFGs and GXMFGs are designed exclusively for graphs with expected average degree going to infinity which, for example, excludes power laws with a coefficient above two.

The learning literature contains various approaches to finding optimal behavior in MFGs and MFC, see \citet{lauriere2022learning} for an overview. For example, \citet{subramanian2022decentralized} develop a decentralized learning algorithm for MFGs where agents independently learn policies, while \citet{guo2019learning, guo2023general} focus on Q-learning methods for general MFGs. Various MFC learning approaches exist \citep{ruthotto2020machine, carmona2023model, gu2023dynamic}, but we are aware of only one work by \citet{hu2023graphon} which learns policies for MFC on dense networks, but not on sparse ones.

Many empirical networks of high practical relevance are considerably sparser than the topologies covered by (LP)GMFGs and GXMFGs. Examples of sparse empirical networks which at least to some extent follow power laws with coefficients between two and three include the internet \citep{vazquez2002large}, coauthorship graphs \citep{goh2002classification} and biological networks \citep{dorogovtsev2002evolution}. These topologies are particularly challenging to analyze \citep{dorogovtsev2002pseudofractal}, especially in combination with particle dynamics \citep{lacker2023local}. Most important in our context, the high fraction of low degree nodes in these networks renders classical mean field approximations and their LPGMFG and GXMFG extensions highly inaccurate.

Both LPGMFGs and GXMFGs assume that the average expected degree diverges to infinity to ensure that the neighborhoods of crucial agents are accurately approximated by a mean field. Since this assumption is inaccurate for various empirical networks, we require a different modeling and learning approach.
To learn policies for such very sparse networks, we employ a suitable graph theoretical convergence principle, \emph{local weak convergence} \citep{van2024random}. Many established graph theoretical models like the configuration model \citep{bollobas1980probabilistic}, the Barab{\'a}si-Albert model \citep{barabasi1999emergence} and the Chung-Lu model \citep{chung2006complex} can model graph sequences converging in the local weak sense. Most importantly in our context, this includes graph sequences with finite average expected degree and power law networks with a coefficient above two, which are neither covered by Lp graphons nor graphexes.

Leveraging local weak convergence, we formulate our new local weak mean field control (LWMFC) model. LWMFC provides a theoretically motivated framework for learning agent behavior in challenging large empirical networks where the average expected degree is finite, but the degree variance may diverge to infinity. On the algorithmic side, we provide a \emph{two systems approximation} for LWMFC and corresponding learning algorithms to approximately learn optimal behaviour in these complex agent networks. Finally, we evaluate our novel LWMFC learning approach for multiple problems on synthetic and real-world networks and compare it to different existing methods mentioned above.
Overall, our contributions can be summarized as:
\begin{itemize}
    \item We introduce LWMFC to model large cooperative agent populations on very sparse graphs with finite expected average degree;
    \item We give a rigorous theoretical analysis and motivation for LWMFC;
    \item We provide a two systems approximation and scalable learning algorithms for LWMFC;
    \item We show the capabilities of our LWMFC learning approach on synthetic and real world networks for different examplary problems.
\end{itemize}

\section{Locally Weak Converging Graphs}

In the following, let $(G_N)_{N \in \mathbb{N}} = (V_N, E_N)_{N \in \mathbb{N}}$ be a growing sequence of random graphs where $V_N$ denotes the vertex set and $E_N$ is the edge set of the corresponding graph $G_N$.
In this paper, we focus on growing graph sequences where the expected average degree remains finite in the limit while the degree variance may diverge to infinity. To formalize the properties of the graph sequences we are focusing on, we first require a suitable graph convergence concept. We choose \textit{local weak convergence in probability} which means that local node neighborhoods converge to neighborhoods in a limiting model. The definition below states local weak convergence, for details see e.g. \citet{lacker2023local}. 

\begin{definition}[Local weak convergence in probability]
A sequence of finite graphs $(G_N)_N$ converges in probability in the local weak sense to a random element $G$ of $\mathcal{G}^*$ if for all continuous and bounded functions $f: \mathcal{G}^* \to \mathbb{R}$
\begin{align*}
    \lim_{N \to \infty} \frac{1}{N} \sum_{i \in [N]} f (C_{v_i} (G_N)) = \mathbb{E} [f (G)] \quad \textrm{in probability}
\end{align*}
where $C_{v_i} (G_N)$ denotes the connected component of $v_i \in G_N$ with root $v_i$ and $\mathcal{G}^*$ is the set of isomorphism classes of connected rooted graphs.
\end{definition}

To obtain meaningful theoretical results and hence practical approximations for large graphs in the next sections, we focus on graph sequences converging in the local weak sense, which we formalize with the next assumption.
\begin{assumption} \label{as:weak_conv}
    The sequence $(G_N)_{N \in \mathbb{N}}$ converges in probability in the local weak sense to some random element $G$ of $\mathcal{G}^*$.
\end{assumption}
The class of random graph sequences fulfilling Assumption \ref{as:weak_conv} covers many famous graph theoretical frameworks which are frequently used in the literature. We will briefly discuss three particularly important types of these models, namely configuration models, preferential attachment models and Chung-Lu graphs. We point to \citet{van2024random} for an extensive introduction and theoretical details and for more random graph models converging in the local weak sense. 

\paragraph{Configuration models.} 
The configuration model \cite{bender1978asymptotic, bollobas1980probabilistic, molloy1995critical, molloy1998size} (CM) is arguably one of the most established random graph models. The basic mechanism of the CM is to start with a fixed and arbitrary degree sequence. Then, a multigraph is randomly generated with the prescribed degree distribution which means that the graph can contain self-edges and double edges between pairs of nodes.

The CM is known to converge under suitable and moderate assumptions in the local weak sense in probability \citep[Theorem 4.1]{van2024random}. However, the CM generates multigraphs instead of simple graphs and the number of multiedges increases drastically as the vertex degrees increase \citep{bollobas1998random}. Consequently, the CM is suboptimal for generating simple graphs with a significant fraction of high degree nodes such as power law networks. 

\paragraph{Preferential attachment models.} 
To model random graphs with power law features, \citet{barabasi1999emergence} introduced the famous Barab{\'a}si-Albert (BA) model. The original BA model was subsequently extended in various ways for different applications, see \citet{piva2021networks} for an overview. The BA model depicts graphs that grow over time by adding new nodes and edges to the topology. Since new nodes are more likely to be connected to highly connected nodes in the current graph, these models are often referred to as \emph{preferential attachment} models.

Preferential attachment models under suitable conditions converge in the local weak sense in probability, see \citet[Theorem 5.8]{van2024random} for details. The BA model, for example, generates power law networks with a coefficient of exactly three \citep{bollobas2001degree}. In many applications, however, it is beneficial to also consider graphs with power law coefficient deviating from three to capture different empirical graph topologies.

\paragraph{Chung-Lu graphs.}
The Chung-Lu (CL) random graph model \citep{aiello2000random, aiello2001random, chung2002connected, chung2006complex} provides an efficient way to model large, sparse networks \citep{fasino2021generating}.
To generate a random CL graph with $N \in \mathbb{N}$ nodes, first specify a weight vector $\boldsymbol{w} \in \mathbb{R}_+^N$ with one weight $w_i \in \mathbb{R}_+$ for each node $i \in \{1, \ldots, N \}$. Then, two nodes $i$ and $j$ are connected with probability $w_i \cdot w_j / \bar{w}$, independently of all other node pairs and with normalization factor $\bar{w} \coloneqq \sum_{1 \leq k \leq N} w_k$. Intuitively, a node with high weight is more likely to have many connections than a node with small weight.

Sequences of CL graphs fulfill Assumption \ref{as:weak_conv} under mild technical assumptions, see \citet[Theorem 3.18]{van2024random} for a formal statement. Most importantly, the average expected degree has to converge to a finite limit, which perfectly aligns with our goal to model very sparse networks.
Throughout the paper, we use the running example of power law degree distributions with coefficient $\gamma >2$ observed in many real world networks to some extent \citep{newman2003structure, kaufmann2009modeling, newman2011structure}. 
\begin{example}[Power law]
    In our work, a power law is a zeta distribution with parameter $\gamma > 2$ such that $P ( \deg (v) = k ) = k^{- \gamma}/\zeta (\gamma)$, where $\zeta (\gamma)$ is the Riemann zeta function $\zeta (\gamma) \coloneqq \sum_{j=1}^\infty j^{- \gamma}$. A power law degree distribution has a finite expectation $\mathbb{E} [\deg (v)] = \zeta (\gamma - 1) / \zeta (\gamma)$ for $\gamma > 2$.
\end{example}
Large, sparse power law networks of the above form can be efficiently generated by the CL framework \citep{fasino2021generating}. Note that our methods apply to all distributions meeting Assumption \ref{as:weak_conv} and perform well on many empirical networks, as shown in the next sections.

\begin{figure*}[ht]
  \centering
  \includegraphics[width=0.99\linewidth]{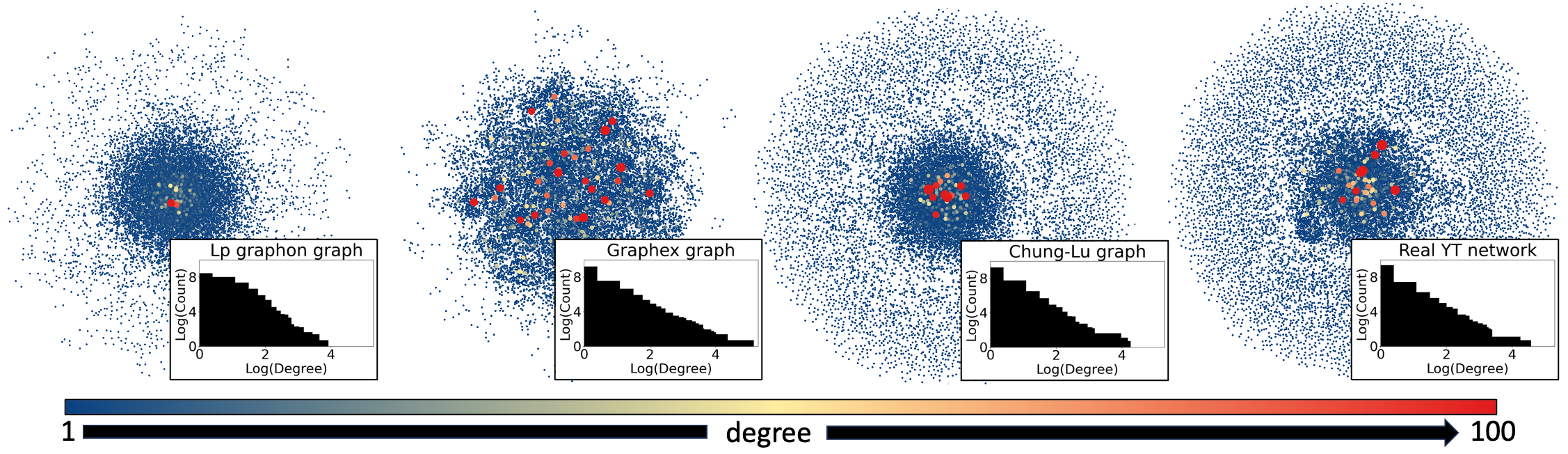}
  \caption{Four networks, first two generated by an Lp graphon and graphex, third is a CL graph and fourth is a real subsampled YouTube (YT) network \citep{mislove2009online, kunegis2013konect}, highly connected nodes are depicted larger. Each network has around 14.5k nodes and 13k edges, except graphex has around 16.5k edges; all networks are plotted in the \emph{prefuse force directed layout} (software: cytoscape). While the Lp graphon graph lacks sufficiently many high degree nodes, the tail of the graphex degree distribution is too heavy. In contrast, the CL graph is qualitatively close to the real YT network.}
  \label{fig:sample_comparison}
\end{figure*}

\paragraph{Advantages over graphons and graphexes.}
Local weak converging graph sequences such as those generated by CM, BM or CL can model sparser, and thus often more realistic topologies than those captured by Lp graphons and graphexes. Figure \ref{fig:sample_comparison} provides an illustration of how Lp graphons, graphexes, and CL graphs compare to a real world subsampled YouTube network \citep{mislove2009online, kunegis2013konect}. Mathematically, both Lp graphons and graphexes are limited to graph sequences were the average degree diverges to infinity. Locally weak converging graph sequences, on the other hand, can capture sparse and often more realistic topologies. The usefulness of models like the CM, BA model and CL graphs is reflected in their frequent use in various research areas. However, formulating a mean field approach based on these graph theoretical models is challenging due to their high number of low degree nodes.

\section{The Finite Model and Its Limit} \label{sec:two}

Denote by $\mathcal{P} (\mathcal{X})$ the set of probability distributions over a finite set $\mathcal{X}$ and define $[N] \coloneqq \{ 1, \ldots, N \}$ for any $N \in \mathbb{N}$.

\paragraph{Finite model.}
Assume some finite state space $\mathcal{X}$, finite action space $\mathcal{U}$ and finite and discrete time horizon $\mathcal{T} \coloneqq \{ 0, \ldots, T - 1 \}$ with terminal time point $T$ are given. Furthermore, there are $N \in \mathbb{N}$ agents connected by some graph $G_N = (V_N, E_N)$ with vertex set $V_N$ and edge set $E_N$. Here, the random state of agent $i \in [N]$ at time $t \in \mathcal{T}$ is denoted by $X_{i,t}^N$. All agents $V^k_N \subseteq V_N$ with degree $k \in \mathbb{N}$ share a common policy $\pi_t^k$ at all time points $t \in \mathcal{T}$. The empirical $k$-degree MF is defined as
\begin{align*}
    \mu_{t}^{N, k} \coloneqq \frac{1}{\vert V_N^k \vert} \sum_{i \in [N]: v_i \in V_N^k} \delta_{X_{i, t}^{N}} \in \mathcal{P} (\mathcal{X})
\end{align*}
for all $t \in \mathcal{T}$ and $k \in \mathbb{N}$.
Define the overall empirical MF sequence as $\mu_t^{N} \coloneqq ( \mu_t^{N, 1}, \mu_t^{N, 2}, \ldots) \in \mathcal{P} (\mathcal{X})^{\mathbb{N}}$.
Each policy $\pi^k \in \mathcal{P} (\mathcal{U})^{\mathcal{T} \times \mathcal{X} \times \boldsymbol{\mathcal{G}}^k}$ in the policy ensemble $\pi = (\pi^1, \pi^2, \ldots) \in \mathcal{P} (\mathcal{U})^{\mathcal{T} \times \mathcal{X} \times \boldsymbol{\mathcal{G}}^k \times \mathbb{N}}$ takes into account the current state of the respective agent $i$ with $k$ neighbors and its neighborhood $\mathbb{G}_{i, t}^{N} \in \boldsymbol{\mathcal{G}}^k \coloneqq \{ G \in \mathcal{P} (\mathcal{X}): k \cdot G \in \mathbb{N}_0^k  \}$. Our learning algorithms also apply to other policy types, e.g., in our experiments we consider computationally efficient policies only depending on the current agent state. Then, the model dynamics are
\begin{align*}
    U_{i, t}^{N} \sim \pi_t^k \left(\cdot \vert X_{i,t}^{N}, \mathbb{G}_{i, t}^{N} \right)
\end{align*}
and
\begin{align*}
    X_{i, t+1}^{N} \sim P \left( \cdot \vert X_{i,t}^{N}, U_{i, t}^{N}, \mathbb{G}_{i, t}^{N} \right)
\end{align*}
for an agent $i$ with degree $k$, $t \in \mathcal{T}$, i.i.d. initial distribution $\mu_0 \in \mathcal{P} (\mathcal{X})$, and transition kernel $P: \mathcal{X} \times \mathcal{U} \times \mathcal{P} \left( \mathcal{X} \right) \to \mathcal{P} \left( \mathcal{X} \right)$. Note that the theory and subsequent learning algorithms extend to degree dependent transition kernels $P^k$.
The policies are chosen to maximize the common objective
\begin{align*}
    J^N(\pi) \coloneqq \sum_{t=1}^T r (\mu^{N}_t)
\end{align*}
with reward function $r: \mathcal{P} (\mathcal{X})^{\mathbb{N}} \mapsto \mathbb{R}$.
Our model also covers reward functions with actions as inputs by using an extended state space $\mathcal{X} \cup (\mathcal{X} \times \mathcal{U})$ and splitting each time step $t \in \mathcal{T}$ into two.

\paragraph{Limiting LWMFC system.}
In the limiting LWMFC system, the MF for each degree $k \in \mathbb{N}$ evolves according to
\begin{align*}
    \mu_{t+1}^{k} &\coloneqq \mu_{t}^{k} P_{t, \boldsymbol{\mu}', W}^{\pi, k} \\
    &\coloneqq \sum_{x \in \mathcal X} \mu^{k}_{t} (x) \sum_{G \in \boldsymbol{\mathcal{G}}^k} P_{\boldsymbol{\pi}} \left( \mathbb{G}_{t}^{k} \left( \boldsymbol{\mu}'_{t} \right) = G \mid x_{t} = x \right) \\
    &\qquad \qquad \qquad \quad \cdot \sum_{u \in \mathcal U} \pi_{t}^k \left(u \mid x, G \right) P \left( \cdot \, \mid x, u, G \right)
\end{align*}
with i.i.d. initial distribution $\mu_0^k \in \mathcal{P} (\mathcal{X})$ and where $\boldsymbol{\mathcal{G}}^k$ is the set of $k$-neighborhood distributions as before.
As in the finite system, define the limiting MF ensemble $\mu_t \coloneqq ( \mu_t^{1}, \mu_t^{2}, \ldots) \in \mathcal{P} (\mathcal{X})^{\mathbb{N}}$
and the corresponding reward in the limiting system is $J (\pi) \coloneqq \sum_{t=1}^T r (\mu_t)$.

\paragraph{Theoretical results.}
Next, we show the strong theoretical connection between the finite and limiting LWMFC system.
The following theoretical results built on the assumption that the underlying graph sequence converges in the local weak sense, formalized by Assumption \ref{as:weak_conv}.
The proofs are in Appendix \ref{app:full_proofs}. We first state empirical MF convergence to the limiting MFs.
\begin{theorem}[MF convergence]\label{thm:mf_conv}
    Under Assumption \ref{as:weak_conv}, for any fixed policy ensemble $\pi$, the empirical MFs converge to the limiting MFs such that for all $k \in \mathbb{N}$ and all $t \in \mathcal T$
    \begin{align*}
        \mu_t^{N, k} \to \mu_t^k \quad \textrm{in probability for} \quad N \to \infty \, .
    \end{align*}
\end{theorem}
The MF convergence from Theorem \ref{thm:mf_conv} enables us to derive a corresponding convergence result for the objective function under a standard continuity assumption on the reward.
\begin{assumption}\label{as:cont_reward}
    The reward function $r: \mathcal{P} (\mathcal{X})^{\mathbb{N}} \mapsto \mathbb{R}$ is continuous.
\end{assumption}
With the above assumption in place, we establish the convergence of the objective function in the finite system to the one in the limiting LWMFC model.
\begin{proposition}[Objective convergence]\label{prop:obj_conv}
    Under Assumptions \ref{as:weak_conv} and \ref{as:cont_reward} and for any fixed policy ensemble $\pi$, the common objective in the finite system converges to the limiting objective, i.e.
    \begin{align*}
        J^{N} (\pi) \to J (\pi) \quad \textrm{in probability for} \quad N \to \infty \, .
    \end{align*}
\end{proposition}
We leverage these findings to show that for a finite set of policy ensembles, the optimal policy for the limiting system in the set is also optimal in all sufficiently large finite systems. Therefore, if one wants to know the optimal ensemble policy for an arbitrary, large agent system, it suffices to find the optimal ensemble policy in the limiting system once which is formalized by Corollary \ref{cor:pol_opt}.
\begin{corollary}[Optimal policy]\label{cor:pol_opt}
    Assume some set $\{ \pi_1, \ldots, \pi_M \}$ of $M < \infty$ policy ensembles is given and that w.l.o.g. $J (\pi_1) > J (\pi_i)$ for all $i \in [M]$ with $i \neq 1$. Under Assumptions \ref{as:weak_conv} and \ref{as:cont_reward} and for some $N^* \in \mathbb{N}$, $\pi_1$ is optimal in all finite systems of size $N > N^*$ such that
    \begin{align*}
        J^{N} (\pi_1) > \max_{i \in [M], i \neq 1} J^{N} (\pi_i) \, .
    \end{align*}
\end{corollary}

\section{The Two Systems Approximation}

In limiting systems on sparse graphs, the state evolution and optimal policy of an agent potentially depend on the entire network \citep{lacker2022case}.
Calculating $P_{\boldsymbol{\pi}} \left( \mathbb{G}_{t}^{k} \left( \boldsymbol{\mu}'_{t} \right) = G \mid x_{t} = x \right)$ at time $t \in \mathcal{T}$ in the limiting system requires all possible $t$-hop neighborhood degree-state distributions where $t$-hop neighborhoods include all agents with a distance of at most $t$ edges to the initial agent. Unfortunately, by Lemma \ref{lem:num_neigh} the number of $t$-hop neighborhoods grows at least exponentially with the degree $k$ in important classes of locally weak converging graphs sequences, such as CL graphs with power laws above two.
\begin{lemma}
    In the limiting system, the number of possible $t$-hop degree-state  neighborhood distributions of agents with degree $k \in \mathbb{N}$ at time $t \in \mathcal{T}$ in the worst case, e.g. CL power law, is $\Omega \left( 2^{\mathrm{poly} (k)} \right)$.\label{lem:num_neigh}
\end{lemma}
Just neglecting high degree nodes in the model might appear as a reasonable approximation to reduce computational complexity. However, the heavy tail of a degree distribution with finite expectation and infinite variance makes this approach highly inaccurate, as Example \ref{examp:high_deg_nodes} illustrates.
\begin{example}
    In a power law graph with $\gamma = 2.5$, around $96 \%$ of nodes have a degree of at most five. However, these $96 \%$ of low degree nodes only account for roughly two thirds of the expected degree, formally $\sum_{h =1}^{5} h^{1 - \gamma} / \zeta (\gamma -1) < 0.68$.
    Nodes with a degree of at most ten still only account for around $76 \%$ of the expected degree.
    \label{examp:high_deg_nodes}
\end{example}

\paragraph{Two systems approximation.}
For the subsequent two systems approximation, we first require a heuristic on the neighbor degree distribution for a given node.
\begin{heuristic} \label{heu:neighbor}
    For an arbitrary node $v' \in V$ the degree distribution of its neighbor $v \in V$ is approximately
    \begin{align*}
        &P (\deg (v) = k \mid \deg (v')= k', (v', v) \in E) \\
        &\qquad \qquad \qquad \qquad \quad \approx \frac{k \cdot P (\deg (v) = k)}{\sum_{k'' \in \mathbb{N}} k'' \cdot P (\deg (v) = k'') }.
    \end{align*}
\end{heuristic}
Heuristic \ref{heu:neighbor} is a good approximation for some sequences of locally weak converging graphs,  such as CL graphs \citep[Chapter 4]{jackson2008social}, and thus reasonable in our setup.
The idea of Heuristic \ref{heu:neighbor} is the following: if one fixes any node $v' \in V$ and considers its neighbors, high degree nodes are more likely to be connected to $v'$ than lowly connected ones. Instead of the overall degree distribution, we thus weight each probability by its degree and normalize accordingly. The result is an approximate neighbor degree distribution accounting for the increased probability of highly connected neighbors.

To address the complexity of the limiting system, we provide an approximate limiting system based on Heuristic \ref{heu:neighbor} and the underlying sparse graph structure. Our two systems approximation consists of a system for small degree agents with at most $k^*$ neighbors and another one for agents with more than $k^*$ connections, where $k^* \in \mathbb{N}$ is some arbitrary, but fixed finite threshold. Define an approximate MF $\hat \mu^k$ for each $k \in [k^*]$ and furthermore summarize all agents with more than $k^*$ connections into the infinite approximate MF $\hat \mu^\infty$ and define $ \hat{\boldsymbol{\mu}} \coloneqq (\hat{\mu}^1, \ldots, \hat{\mu}^{k^*}, \hat{\mu}^\infty)$. Based on Heuristic \ref{heu:neighbor}, we assume that all agents with more than $k^*$ neighbors observe the same neighborhood state distribution
\begin{align*}
    \hat{\mathbb{G}}_{t}^{\infty} (\hat{\boldsymbol{\mu}}) &\coloneqq \frac{1}{\mathbb{E} [\deg (v)]} \left( \sum_{k = k^* + 1}^{\infty}  k P (\deg (v) = k) \right) \hat{\mu}_t^\infty \\
    &\qquad + \frac{1}{\mathbb{E} [\deg (v)]} \sum_{h =1}^{k^*} h P (\deg (v) = h)  \hat{\mu}_t^h \, .
\end{align*}
The unified approximate neighborhood state distribution $\hat{\mathbb{G}}_{t}^{\infty}$ allows us to state an approximate, simplified version of the MF forward dynamics for high degree agents given by
\begin{align*}
    &\hat{\mu}_{t+1}^{\infty} \coloneqq \hat{\mu}_{t}^{\infty} \hat{P}_{t, \boldsymbol{\mu}', W}^{\pi, \infty}
    \coloneqq \\
    &\, \sum_{x,u} \hat{\mu}^{\infty}_{t} (x)  \pi_{t}^\infty \left(u \mid x, \hat{\mathbb{G}}_{t}^{\infty} (\boldsymbol{\mu}') \right) P \left(\cdot \, \mid x, u, \hat{\mathbb{G}}_{t}^{\infty} (\boldsymbol{\mu}') \right)
\end{align*}
where all agents with more than $k^*$ connections follow the same policy $\pi_{t}^\infty \in \mathcal{P} (\mathcal{U})^{\mathcal{T} \times \mathcal{X} \times \mathcal{P} (\mathcal{X})}$ and where the sum is over all $(x,u) \in \mathcal{X} \times \mathcal{U}$.
The approximate neighborhood of an agent with degree $k \in [k^*]$ at each time $t \in \mathcal{T}$ is sampled from $\hat{\mathbb{G}}_{t}^{k} (\hat{\boldsymbol{\mu}}) \sim \mathrm{Mult} (k, \hat{\mathbb{G}}_{t}^{\infty} (\hat{\boldsymbol{\mu}}))$, i.e. $\hat{\mathbb{G}}_{t}^{k} (\hat{\boldsymbol{\mu}})$ is multinomial with $k$ trials and probabilities $\hat{\mathbb{G}}_{t}^{\infty} (\hat{\boldsymbol{\mu}}) (x)$ for each $x \in \mathcal{X}$.
Using Heuristic \ref{heu:neighbor}, the approximation yields for each $k \in [k^*]$ the MF forward dynamics
\begin{align*}
    &\hat{\mu}_{t+1}^{k} \coloneqq \hat{\mu}_{t}^{k} \hat{P}_{t, \boldsymbol{\mu}', W}^{\pi, k} \coloneqq \\
    &\sum_{x,u,G} \hat{\mu}^{k}_{t} (x) P_{\mathrm{Mult}} \left( \hat{\mathbb{G}}_{t}^{k} = G \right) \pi_{t}^k \left(u \mid x, G \right) P \left(\cdot \, \mid x, u, G \right)
\end{align*}
where the sum is over all $(x,u, G) \in \mathcal{X} \times \mathcal{U} \times \boldsymbol{\mathcal{G}}^k$.

\paragraph{Extensive approximation.}
In Appendix \ref{app:ext_approx_deriv}  we derive a second, extensive approximation 
\begin{align*}
    &P_{\boldsymbol{\pi}, \boldsymbol{\mu}} \left( \mathbb{G}_{t + 1}^{k}  \left( {\boldsymbol \mu}_{t} \right) = G, x_{t+1} = x \right) \\
    &\quad\approx \sum_{G' \in \boldsymbol{\mathcal{G}}^k} \sum_{x' \in \mathcal X} \sum_{c \in \boldsymbol{\mathcal{C}}^k} P_{\boldsymbol{\pi}, \boldsymbol{\mu}} \left( \mathbb{G}_{t}^{k}  \left( {\boldsymbol \mu}_{t} \right) = G', x_t = x' \right) \\ 
    &\qquad\qquad\qquad\qquad \cdot \left[ \sum_{u \in \mathcal{U}} \pi^k \left( u \mid x' \right) P \left( x \mid x', u, G'\right) \right] \\
    &\qquad\qquad\qquad\qquad \cdot \left[ \sum_{\boldsymbol{a}_2 \in \boldsymbol{\mathcal{A}}^k_2 (G', c)} \prod_{j} \mathrm{Mult}_{\boldsymbol{p}_{2, j}} (\boldsymbol{a}_{2, j}) \right] \\
    &\qquad \qquad \quad \cdot \frac{ \sum_{\boldsymbol{a}_3 \in \boldsymbol{\mathcal{A}}^k_3 (G, G', c)} \prod_{j, m} \mathrm{Mult}_{\boldsymbol{p}_{3, j m}} (\boldsymbol{a}_{3, j m})}
    {\sum_{\boldsymbol{a}_2 \in \boldsymbol{\mathcal{A}}^k_2 (G', c)} \prod_{j, m} p_{j,m} (\boldsymbol{a}_2)} \, .
\end{align*}
of the finite agent neighborhoods in the LWMFC model where we use the abbreviated notation $p_{j,m} (\boldsymbol{a}_2) \coloneqq \left( P \left(\deg (v) = m \mid (v', v) \in E \right) \mu_t^m (s_j) \right)^{a_{j m}}$. Here, the idea is to go beyond the previous multinomial assumption $\hat{\mathbb{G}}_{t}^{k} (\hat{\boldsymbol{\mu}}) \sim \mathrm{Mult} (k, \hat{\mathbb{G}}_{t}^{\infty} (\hat{\boldsymbol{\mu}}))$ and to use state-degree neighborhood distributions $\boldsymbol{a}_2 \in \boldsymbol{\mathcal{A}}^k_2 (G', c)$ and state-state-degree neighborhood distributions $\boldsymbol{a}_3 \in \boldsymbol{\mathcal{A}}^k_3 (G, G', c)$ to capture agents changing from $x \in \mathcal{X}$ to $x'\in \mathcal{X}$ at a time step. We provide the extensive approximation derivation and corresponding definitions of sets like $\boldsymbol{\mathcal{A}}^k_2 (G', c)$ and $\boldsymbol{\mathcal{A}}^k_3 (G, G', c)$ in Appendix \ref{app:ext_approx_deriv}. As we will see in the following, the extensive approximation often shows a moderately higher accuracy than our first approximation. However, the accuracy boost entails a significantly higher computational complexity due to multiple sums over sets like $\boldsymbol{\mathcal{A}}^k_2 (G', c)$ and $\boldsymbol{\mathcal{A}}^k_3 (G, G', c)$. Thus, our first approximation is more practical since it combines reasonable accuracy with low computational complexity 
while the extensive approximation is computationally too expensive for our purposes.

\section{Learning Algorithms}
To solve the MARL problem of finding optimal policies for each class of $k$-degree nodes, we propose two methods based on reducing the otherwise intractable many-agent graphical system to a single-agent MFC MDP. The first approach in Algorithm~\ref{algo:lwmfc} is based on solving the resulting limiting MFC MDP under the parameters of the real graph, using the previously established two systems approximation. The second approach in Algorithm~\ref{algo:clmfmarl} instead directly learns according to single-agent RL that solves the MFC MDP by interacting with the real graph.

\paragraph{RL in MFC MDP.}
The two system approximation reduces the complexity of otherwise intractable large interacting systems on networks to the MFs of each degree. The system state at any time is then given by low-degree MFs $\mu^1_t, \mu^2_t, \ldots, \mu^{k^*}_t$ and high-degree MF $\mu^\infty_t$, briefly $\boldsymbol \mu_t \coloneqq (\mu^1_t, \mu^2_t, \ldots, \mu^{k^*}_t, \mu^\infty_t)$. Given a state $\boldsymbol \mu_t$, the possible state evolutions depend only on the analogous set of low-degree and high-degree policies at that time, $\boldsymbol \pi_t \coloneqq (\pi^1_t, \pi^2_t, \ldots, \pi^{k^*}_t, \pi^\infty_t)$. Therefore, choosing a $\boldsymbol \pi_t$ fully defines the state transition of the overall system, and is thus considered as the \emph{high-level action} in the MFC MDP. Introducing a high-level policy $\hat \pi$ to output $\boldsymbol \pi_t \sim \hat \pi_t(\boldsymbol \pi_t \mid \boldsymbol \mu_t)$ allows us to solve for an optimal set of policies by solving the MFC MDP for optimal $\hat \pi$, since the limiting MF dynamics are deterministic. Finally, the MFC MDP is solved by applying single-agent policy gradient RL, resulting in Algorithm~\ref{algo:lwmfc}. In practice, we use proximal policy optimization \citep{schulman2017proximal}. To lower the complexity of the resulting MDP, we parametrize policies as distributions over actions given the node state, $\pi^k_t \in \mathcal P(\mathcal U)^{\mathcal X}$.

\begin{algorithm}[ht!]
    \caption{\textbf{LWMFC Policy Gradient}}
    \label{algo:lwmfc}
    \begin{algorithmic}[1]
        \FOR {iterations $n = 1, 2, \ldots$}
            \FOR {time steps $t = 0, \ldots, B_{\mathrm{len}}-1$}
                \STATE Sample LWMFC MDP action $\boldsymbol \pi_t \sim \hat \pi^\theta(\boldsymbol \pi_t \mid \boldsymbol \mu_t)$.
                \STATE Compute reward $r(\boldsymbol \mu_t)$, next MF $\boldsymbol \mu_{t+1}$, termination flag $d_{t+1} \in \{0, 1\}$.
            \ENDFOR
            \STATE Update policy $\hat \pi^\theta$ on minibatches $b \subseteq \{ (\boldsymbol \mu_t, \boldsymbol \pi_t, r_t, d_{t+1}, \boldsymbol \mu_{t+1}) \}_{t \geq 0}$ of length $b_{\mathrm{len}}$.
        \ENDFOR
    \end{algorithmic}
\end{algorithm}

\paragraph{MARL on real networks.}

\begin{table*}[t!]
  \vskip -0.05in
  \caption{Average expected total variation $\Delta \mu = \frac 1 {2T} \mathbb{E} \left[ \sum_t \lVert \hat \mu_t - \mu_t \rVert_1 \right] \in [0, 1]$ of MF $\mu_t$ and empirical MF $\hat \mu_t = \sum_i \delta_{X^i_t}$ ($\pm$ standard deviation, 50 trials), for the four models for four problems on eight real-world networks. Extensive LWMFC* not displayed for last two problems since calculations exceed maximum runtime. Best result for each network-problem combination in bold.}
  \label{table:dynamics_compared}
  \vskip 0.1in
  \begin{center}
  \setlength{\tabcolsep}{4pt} 
  {\fontsize{9pt}{8.4pt}\selectfont
  \renewcommand\arraystretch{1.05}  
  { 
  \begin{tabular}{llcccccccc}
    \toprule
    \multicolumn{2}{c}{\multirow{2}{*}{Model}} & \multicolumn{8}{c}{Average expected total variation $\Delta \mu$ in $\%$, standard deviation in brackets}\\
    \cmidrule(lr){3-10}
     & & CAIDA & Cities & Digg Friends & Enron &  Flixster & Slashdot & Yahoo & YouTube \\
    \midrule
    \multirow{4}*{\rotatebox{90}{SIS}} & LPGMFG & 24.02 (1.25)  & 28.16 (0.41) & 21.98 (0.26) & 24.77 (0.32) & 22.48 (0.07)  & 23.70 (0.43) & 10.11 (2.10) & 22.94 (0.25) \\
    & GXMFG & 9.07 (1.25) & 10.90 (0.41) & 4.72 (0.26) & 4.73 (0.32) & 3.78 (0.07) & 5.48 (0.43) & 9.31 (2.10) & 6.43 (0.25) \\
    & LWMFC & 2.59 (1.14) & 5.00 (0.40) & 3.57 (0.26) & 3.39 (0.31) & 1.60 (0.07)  & 2.41 (0.43) & \textbf{3.59 (1.59)} & 3.53 (0.25)  \\
    & LWMFC* & \textbf{1.75 (0.90)} & \textbf{4.20 (0.40)} & \textbf{3.02 (0.26)} & \textbf{2.67 (0.31)} & \textbf{ 0.90 (0.07)}  & \textbf{1.70 (0.42)} & 3.81 (1.70) & \textbf{2.93 (0.25)}  \\
    \midrule
    \multirow{4}*{\rotatebox{90}{SIR}} & LPGMFG & 9.11 (1.40)  & 10.01 (0.34) & 8.68 (0.31) & 9.51 (0.32) & 8.99 (0.09)  & 9.37 (0.38) & 4.88 (1.82) & 8.90 (0.23) \\
    & GXMFG & 2.81 (1.10)  & 2.63 (0.31) & 1.27 (0.29) & 0.99 (0.30) & 0.99 (0.09) & 1.58 (0.36) & 4.60 (1.71) & 1.79 (0.23) \\
    & LWMFC & 1.31 (0.87)  & 1.36 (0.27) & 1.08 (0.28) & 0.91 (0.30) & 0.58 (0.08)  & 0.99 (0.33) & \textbf{2.62 (1.30)} & 1.07 (0.23)  \\
    & LWMFC* & \textbf{1.18 (0.82)}  & \textbf{1.10 (0.26)} & \textbf{0.80 (0.27)} & \textbf{0.59 (0.28)} & \textbf{0.26 (0.08)} & \textbf{0.71 (0.29)} & 2.63 (1.30) & \textbf{0.78 (0.23)}  \\
    \midrule
    \multirow{3}*{\rotatebox{90}{Color}} & LPGMFG & 38.73 (0.17)  & 38.59 (0.09) & 38.70 (0.04) & 39.83 (0.06) & 39.55 (0.02)  & 39.07 (0.06) & 34.18 (0.26) & 38.52 (0.04) \\
    & GXMFG & 11.33 (0.13) & 7.90 (0.06) & 7.85 (0.02) & 4.91 (0.03) & 6.38 (0.01) & 6.81 (0.03) & 32.62 (0.24) & 8.76 (0.02) \\
    & LWMFC & \textbf{0.70 (0.12)}  & \textbf{0.48 (0.05)} & \textbf{0.19 (0.02)} & \textbf{0.36 (0.04)} & \textbf{0.39 (0.02)} & \textbf{0.33 (0.04)} & \textbf{1.05 (0.19)} & \textbf{0.19 (0.03)} \\
    \midrule
    \multirow{3}*{\rotatebox{90}{Rumor}} & LPGMFG & 20.03 (2.15)  & 22.56 (0.50) & 18.39 (0.55) & 20.27 (0.61) & 18.94 (0.16)  & 19.70 (0.82) & 9.68 (3.76) & 19.23 (0.47) \\
    & GXMFG & 6.98 (2.06)  & 7.49 (0.49) & 3.33 (0.54) & 2.86 (0.58) & 2.65 (0.16) & 3.82 (0.79) & 9.01 (3.69) & 4.79 (0.47) \\
    & LWMFC & \textbf{3.06 (1.59)}  & \textbf{4.31 (0.48)} & \textbf{ 3.00 (0.53)} & \textbf{ 2.62 (0.57)} & \textbf{ 1.73 (0.15)}  & \textbf{2.41 (0.75)} & \textbf{5.01 (2.21)} & \textbf{3.27 (0.46)}  \\
    \bottomrule
  \end{tabular}}
  }
  \end{center}
  \vskip -0.1in
\end{table*}

In addition to assuming knowledge of the model and computing the limiting MFC MDP equations, we may also directly learn on real network data without such model knowledge in a MARL manner. To do so, we still apply policy gradient RL to solve an assumed MFC MDP, but substitute samples from the real network into $\boldsymbol \mu_t$. At the same time, we let each node perform its actions according to the sampled $\boldsymbol \pi_t \sim \hat \pi_t(\boldsymbol \pi_t \mid \boldsymbol \mu_t)$. This approach is well justified by the previous theory and approximation, as for sufficiently large networks the limiting system and therefore also its limiting policy gradients are well approximated by this procedure. 

\begin{algorithm}[h!]
    \caption{\textbf{LWMFMARL Policy Gradient}}
    \label{algo:clmfmarl}
    \begin{algorithmic}[1]
        \FOR {iterations $n = 1, 2, \ldots$}
            \FOR {time steps $t = 0, \ldots, B_{\mathrm{len}}-1$}
                \STATE Sample LWMFC MDP action $\boldsymbol \pi_t \sim \hat \pi^\theta(\boldsymbol \pi_t \mid \boldsymbol \mu_t)$.
                \FOR {node $i = 1, \ldots, N$}
                    \STATE Sample per-node action $U_{i,t} \sim \pi^{k_i}_t(U_{i,t} \mid X_{i,t})$ with degree $k_i = \infty$ if $k_i > k^*$.
                \ENDFOR
                \STATE Perform actions, observe reward $r_t$, next MF $\boldsymbol \mu_{t+1}$, termination flag $d_{t+1} \in \{0, 1\}$.
            \ENDFOR
            \STATE Update policy $\hat \pi^\theta$ on minibatches $b \subseteq \{ (\boldsymbol \mu_t, \boldsymbol \pi_t, r_t, d_{t+1}, \boldsymbol \mu_{t+1}) \}_{t \geq 0}$ of length $b_{\mathrm{len}}$.
        \ENDFOR
    \end{algorithmic}
\end{algorithm}

The approach results in Algorithm~\ref{algo:clmfmarl} and has advantages. Firstly, the algorithm does not assume model knowledge and is therefore a true MARL algorithm, in contrast to solving the limiting MFC MDP. Secondly, the algorithm avoids potential inaccuracies of the two systems approximation, as we will see in Section~\ref{sec:exp}, since it directly interacts with a real network of interest. Lastly, in contrast to standard independent and joint learning MARL methods, the method is rigorously justified by single-agent RL theory and avoids exponential complexity in the number of agents respectively.

\section{Examples}
For a general overview of many applications, see \citet{lauriere2022learning}.
We consider four problems briefly described here. Problem details can be found in Appendix \ref{sec:model_details}.

\begin{figure}[b!]
	\centering   
    \includegraphics[width=0.99\linewidth]{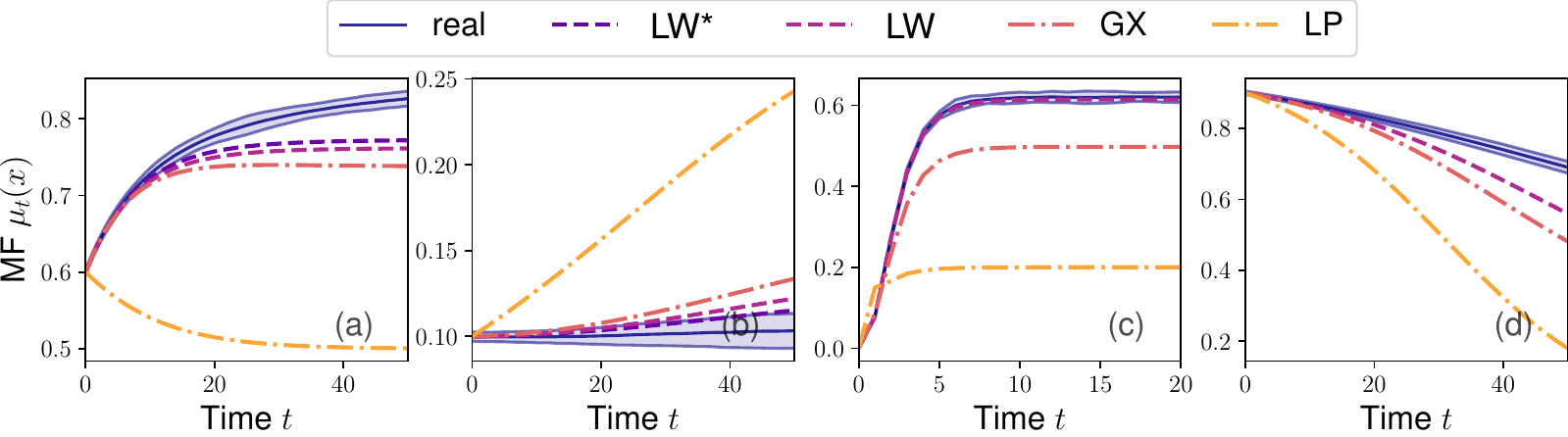}
	\caption{Overall MF evolution on real networks (50 trials, with two std. devs.), for our approx. (LW), our extensive approx. (LW$^*$), graphex (GX), and Lp graphon (LP) models: (a) SIS on Enron, (b) SIR on Slashdot, (c) Color on CAIDA (without LW$^*$), (d) Rumor on Cities (without LW$^*$).}
 \label{fig:dynamics_time_comparison}
\end{figure}

\paragraph{Susceptible-Infected-Susceptible/Recovered (SIS/SIR).}
The classical SIS model \citep{kermack1927contribution, brauer2005kermack} is a benchmark in the MF learning literature \citep{lauriere2022scalable, zhou2024graphon}. Agents are infected or susceptible, resulting in the state space $\mathcal{X} \coloneqq \{ S, I \}$, and decide to protect themselves or not. The infection probability increases without protection, and with the number of infected neighbors.
Furthermore, there is a constant probability for recovering from an infection.
The SIR model \citep{hethcote2000mathematics, doncel2022mean} is an extension of SIS where agents can also be in a recovered state $R$ where they are immune to a reinfection. Consequently, the state space for the SIR model is $\mathcal{X} \coloneqq \{ S, I, R \}$.

\paragraph{Graph coloring (Color).}
Inspired by graph coloring \citep{jensen2011graph, barenboim2022distributed}, the states are finitely many colors on a circle and a target color distribution is given. Agents stay at their color or costly move to a neighboring color. The objective decreases for deviations from the target color distribution and if neighbors of an agent have neighboring colors to the agent's color on the respective color ring.

\paragraph{Rumor.}
In the rumor model \citep{maki1973mathematical, gani2000maki, cui2022hypergraphon}, agents are either aware of a rumor and are consequently in the aware state $A$ or they have not heard the rumor and are in the ignorant state $I$. Aware agents decide whether they spread the rumor to their neighbors or keep it to themselves. They are awarded for spreading the rumor to unaware agents but loose reputation for telling the rumor to already aware agents.

\begin{table*}[t!]
  \caption{(LW)MFC policy gradient, (LW)MFMARL policy gradient, and IPPO for four problems on synthetic CL graphs of  size $N$. Best objective after 24 hours of training on 96 CPUs. Best result for each problem-graph tuple in bold. }
  \label{tab:learning_algo_comparison}
  \begin{center}
  \setlength{\tabcolsep}{4pt}
  \renewcommand\arraystretch{1.05}  
  \resizebox{0.999\textwidth}{!}{ 
  {\fontsize{7pt}{8.4pt}\selectfont
  \begin{tabular}{lcccccccccccc}
    \toprule
    \multicolumn{1}{c}{\multirow{2}{*}{Problem}} & \multicolumn{3}{c}{$N = 167$} & \multicolumn{3}{c}{$N = 406$} & 
    \multicolumn{3}{c}{$N = 860$} & \multicolumn{3}{c}{$N = 1598$} \\
    \cmidrule(lr){2-4} \cmidrule(lr){5-7} \cmidrule(lr){8-10} \cmidrule(lr){11-13}
    & IPPO & MFC & MFMARL & IPPO & MFC & MFMARL & IPPO & MFC & MFMARL  & IPPO & MFC & MFMARL \\
    \midrule
    SIS & -20.80 & -14.56 & \textbf{-12.50} & -21.40 & -14.18 & \textbf{-11.64} & -19.70 & -12.42 & \textbf{-9.11} & -22.42 & -13.51 & \textbf{-11.13} \\ 
    SIR & -7.45 & -7.84 & \textbf{-6.99} & -7.18 & -7.42 & \textbf{-6.55} & -10.64 & -6.86 & \textbf{-5.15} & -7.73 & -7.42 & \textbf{-6.32} \\
    Color & -8.20 & -6.84 & \textbf{-6.74} & -8.05 & -7.04 & \textbf{-6.98} & -8.48 & -7.08 & \textbf{-5.85} & -8.15 & -6.97 & \textbf{-6.94} \\
    Rumor & 0.24 & \textbf{1.19} & 0.27 & 0.16 & \textbf{1.33} & 0.19 & 0.25 & \textbf{1.47} & 1.35 & 0.12 & \textbf{1.33} & 0.17 \\
    \bottomrule
  \end{tabular}
  }
  }
  \end{center}
\end{table*}

\section{Simulation \& Results} \label{sec:exp}
In this section, we numerically verify the two system approximation as well as the proposed learning algorithms by comparing them with baselines from the literature. The two systems approximation is compared with previous graph approximations such as graphex or Lp graphon MF equations, and the learning algorithms are verified against standard scalable independent learning methods such as IPPO \citep{tan1993multi, papoudakis1benchmarking}, due to the large scale of networks considered here.
To generate artificial networks of different sizes we employ a CL-based graph sampling algorithm \citep{chung2002connected, miller2011efficient} from the Python NetworkX package.

We compare the accuracy of our model on different empirical datasets with Lp graphon and graphex based models and with our extensive approximation LWMFC$^*$, where computationally feasible, to see how much information is lost in the LWMFC approximation. 
We use eight datasets from the KONECT database \citep{kunegis2013konect}, where  we substitute directed or weighted edges by simple undirected edges:
CAIDA \citep{leskovec2007graph}($N \approx 26k$),
Cities \citep{kunegis2013konect} ($N \approx 14k$),
Digg Friends \citep{hogg2012social} ($N \approx 280k$),
Enron \citep{klimt2004enron} ($N \approx 87k$),
Flixster \cite{Zafarani+Liu:2009} ($N \approx 2.5mm$),
Slashdot \citep{gomez2008statistical} ($N \approx 50k$),
Yahoo \citep{kunegis2013konect} ($N \approx 653k$), and 
YouTube \citep{mislove2009online} ($N \approx 3.2mm$).
See the references for details on the respective empirical networks.

\paragraph{Results.}
First, we establish the usefulness of LWMFC and LWMFC$^*$ by comparing their dynamics to those of LPGMFGs and GXMFGs \citep{fabian2023learning, fabian2024learning} on eight real-world networks, see Figure \ref{fig:dynamics_time_comparison} for examplary dynamics over time. As Table \ref{table:dynamics_compared} shows, our LWMFC approach clearly outperforms the current LPGMFGs and GXMFGs methods for all empirical networks and problems. The extensive approximation LWMFC$^*$ moderately outperforms LWMFC across datasets, except Yahoo. Since the extensive approximation is more detailed, it is often more accurate than the LWMFC approximation. However,  Table \ref{table:dynamics_compared} lacks an evaluation for LWMFC$^*$ on the Color and Rumor problem because the extensive approximation is computationally too expensive for these problems. Consequently, LWMFC dynamics are the more practical choice since they are computationally tractable and yield a good performance across problems and datasets.

\begin{figure}[h!]
	\centering
  \includegraphics[width=0.99\linewidth]{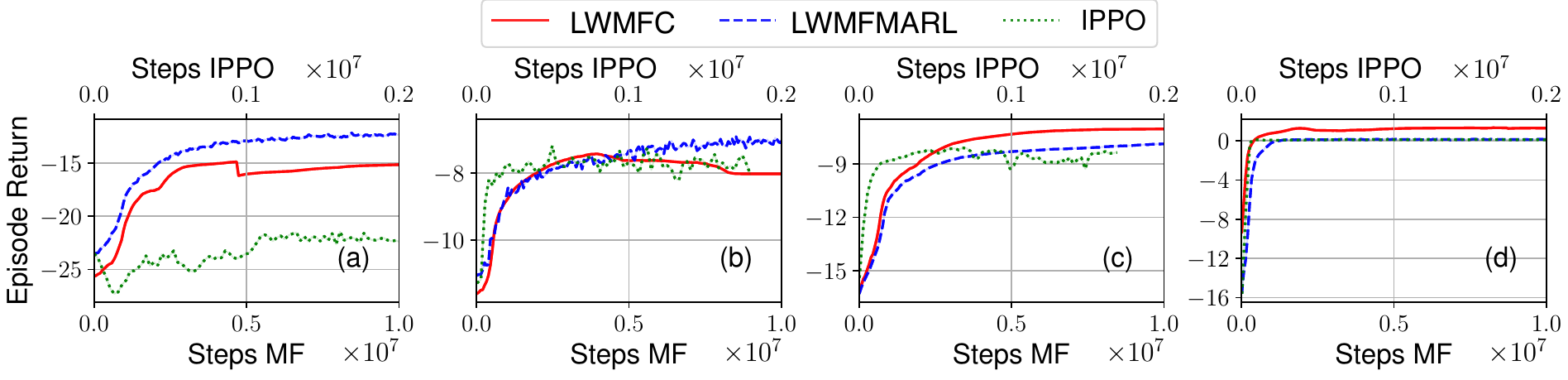}
	\caption{Training curves of LWMFC policy gradient, LWMFMARL, and IPPO on a random CL graph with 406 nodes for: (a) SIS, (b) SIR, (c) Color, (d) Rumor.}
 \label{fig:synth_learning_curves}
\end{figure}

\begin{figure}[h!]
	\centering
  \includegraphics[width=0.99\linewidth]{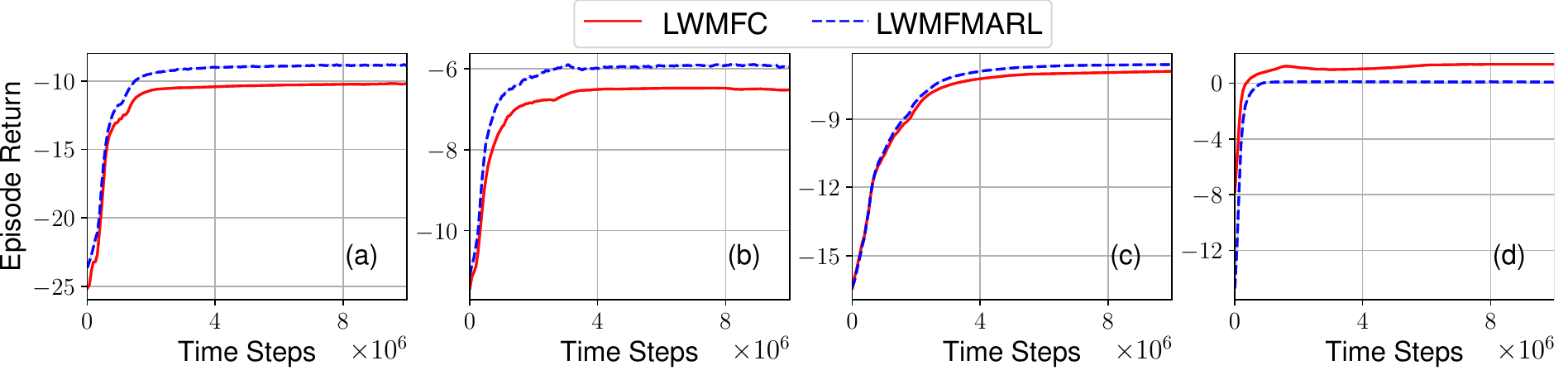}
	\caption{Training curves of LWMFC policy gradient and LWMFMARL for four different examples: (a) SIS on Enron, (b) SIR on Slashdot, (c) Color on CAIDA, (d) Rumor on Cities.}
 \label{fig:real_network_learning_curves}
\end{figure}

The second part of our results focuses on our two learning algorithms LWMFC and LWMFMARL and compares them to the well-known IPPO algorithm. In Table \ref{tab:learning_algo_comparison}, our algorithms outperform IPPO for all problems on the two larger graphs with $860$ and $1598$ nodes, respectively. On the two smaller graphs, LWMFC and LWMFMARL still yield an at least competitive performance compared to IPPO, where IPPO is only marginally better than LWMFC on two problem instances, namely SIR on $N=167$ and $N=406$. 
We point out that LWMFC, in contrast to IPPO and LWMFMARL, is not evaluated on the empirical system, but by design on the limiting LWMFC model, which may differ from the true system behavior. 
These findings are complemented by the corresponding training curves in Figure \ref{fig:synth_learning_curves}. Finally, Figure \ref{fig:real_network_learning_curves} depicts how the training curves of our LWMFC and LWMFMARL algorithms converge on different empirical networks for different problems.

\section{Conclusion}

We have introduced the novel LWMFC framework which can depict agent networks with finite expected degree and diverging variance. After a theoretical analysis, we provided a practical two systems approximation which was then leveraged to design scalable learning algorithms. Finally, we evaluated the performance of our model and learning algorithms for different problems on synthetic and real-world datasets and compared them to existing methods. For future work, one could extend the LWMFC model to various types of specific mean field models, e.g. to partial observability or agents under bounded rationality. We hope that LWMFC and the corresponding learning approaches prove to be a versatile and useful tool for researchers across various applied research areas.

\section*{Acknowledgements}
This work has been co-funded by the Hessian Ministry of Science and the Arts (HMWK) within the projects "The Third Wave of Artificial Intelligence - 3AI" and hessian.AI, and the LOEWE initiative (Hesse, Germany) within the emergenCITY center. The authors acknowledge the Lichtenberg high performance computing cluster of the TU Darmstadt for providing computational facilities for the calculations of this research.

\section*{Impact Statement}
This paper presents work whose goal is to advance the field of machine learning. There are many potential societal consequences  of our work, none which we feel must be specifically highlighted here.

\bibliography{example_paper}
\bibliographystyle{icml2025}


\newpage
\appendix
\onecolumn
\section{Appendix: Proofs for the Theoretical Results} \label{app:full_proofs}

\subsection{Proof of Theorem \ref{thm:mf_conv}}

\begin{proof}
    We aim to eventually apply \citet[Theorem 3.6]{lacker2023local} and therefore have to check that the respective conditions hold in our model.
    Keeping in mind Assumption \ref{as:weak_conv} and the i.i.d. initial distribution $\mu_0$, we leverage \citet[Corollary 2.16]{lacker2023local} to obtain convergence in probability in the local weak sense of the marked graphs $(G_N, X^{G_N})$ to the limiting marked graph $(G, X^G)$.

    Since the theory in \citet{lacker2023local} is only formulated in terms of particle systems without including actions in the form of policies, we provide a suitable reformulation of our LWMFC model. Thus, define an auxiliary extended state space $\mathcal{X}_{\mathrm{e}} \coloneqq \mathcal{X} \cup (\mathcal{X} \times \mathcal{U})$ which serves as the state space for the extended particle system for some fixed policy ensemble $\pi$. The idea behind the extended state space $\mathcal{X}_{\mathrm{e}}$ is to define an extended particle system where the state transition in $\mathcal{X}$ and the choice of the next action $u_{t+1} \in \mathcal{U}$ are separated into two different time steps.

    Using the notations from \citet{lacker2023local}, denote by $\mathcal{S}^{\sqcup} (\mathcal{X})$ the set of finite unordered sequences of arbitrary length with values in $\mathcal{X}$ and by $\Xi\coloneqq \mathcal{X}^{\mathcal{X} \times \mathcal{S}^{\sqcup} (\mathcal{X})} \times \mathcal{U}^{\mathcal{X} \times \mathcal{U} \times \mathcal{S}^{\sqcup} (\mathcal{X})}$ the set of possible noise values.
    Next, specify a transition function $F^\tau: \mathcal{X}_{\mathrm{e}} \times \mathcal{S}^{\sqcup} (\mathcal{X}_{\mathrm{e}}) \times \Xi \to \mathcal{X}_{\mathrm{e}}$ for each $\tau \in \mathcal{T}_{\mathrm{e}} \coloneqq \{0\} \cup [2 T-1]$ by
    \begin{align*}
        X_{\mathrm{e},i,\tau+1}^N = F^\tau (X_{\mathrm{e},i,\tau}^N, \mathbb{G}_{\mathrm{e}, i,\tau}^N, \xi_{i, \tau+1}) \coloneqq
        \begin{cases}
            (X_{\mathrm{e},i,\tau}^N, \xi_{i, \tau+1}^0 (X_{\mathrm{e},i,\tau}^N, \mathbb{G}_{\mathrm{e}, i,\tau}^N) ) \quad \textrm{if} \quad \tau/2 \in \{ 0 \} \cup \mathbb{N}\\
            \phantom{X_{\mathrm{e}}^N} \xi_{i, \tau+1}^1 (X_{\mathrm{e},i,\tau}^N, \mathbb{G}_{\mathrm{e}, i,\tau}^N) \qquad \quad \textrm{otherwise,}
        \end{cases}
    \end{align*}
    where the neighborhood in the extended particle system $\mathbb{G}_{\mathrm{e}, i,\tau}^N$ corresponds to $\mathbb{G}_{i,\lfloor \tau /2 \rfloor}^N$ in the original system.
    Here, the noise terms $\xi_{i, \tau + 1} = (\xi_{i, \tau + 1}^0, \xi_{i, \tau + 1}^1 )$ depict the used noise depending on whether $\tau$ is an even or odd number. If $\tau$ is an even number, i.e. $\tau/2 \in \{ 0 \} \cup \mathbb{N}$,  we use $\xi_{i, \tau + 1}^0 (X_{\mathrm{e},i,\tau}^N, \mathbb{G}_{\mathrm{e}, i,\tau}^N)$ which is a $\mathcal{U}$-valued random variable with distribution
    \begin{align*}
        \xi_{i, \tau + 1}^0 (x, G)
        \sim \pi^k_{\tau /2} \left( \cdot \mid x, G \right)
    \end{align*}
    for each neighborhood $G$ and state $x \in \mathcal X$, where $k$ is the degree of agent $i$. If $\tau$ is odd, i.e. we have $X_{\mathrm{e},i,\tau}^N \in \mathcal{X} \times \mathcal{U}$, we choose the first, $\mathcal{X}$-valued entry of $X_{\mathrm{e},i,\tau}^N$ as the $x$ in the above probability distribution.
    
    The $\mathcal{X}$-valued noise component $\xi_{i, \tau + 1}^1$ is distributed as follows: if $\tau$ is an odd number, the noise term is  sampled from
    \begin{align*}
        \xi_{i, \tau + 1}^1 (x, u, G)
        \sim P \left( \cdot \vert x,u,G \right)
    \end{align*}
    where $X_{\mathrm{e},i,\tau}^N = (x,u)$. If $\tau$ is even and thus $X_{\mathrm{e},i,\tau}^N \in \mathcal X$, we just choose some arbitrary, but fixed action $u' \in \mathcal{U}$ instead of $u$ in the above sampling process.
    
    Now, it remains to check that \citet[Assumption A]{lacker2023local} is satisfied by the extended particle system defined above. First, the noise terms $\xi_{i, \tau}$ are i.i.d. distributed for all agents $i \in [N]$ and with respect to all time points $\tau \in \mathcal{T}_{\mathrm{e}}$ by construction. 
    Finally, keeping in mind that the respective spaces are discrete, the map $F^\tau$ is continuous for each $\tau \in \mathcal{T}_{\mathrm{e}}$. Therefore, \citet[Theorem 3.6]{lacker2023local} yields the desired result.
\end{proof}

\subsection{Proof of Proposition \ref{prop:obj_conv}}

\begin{proof}
    We want to show 
    \begin{align*}
        J^{N} (\pi) \to J (\pi) \quad \textrm{in probability for} \quad N \to \infty
    \end{align*}
    which is equivalent to
    \begin{align*}
        \sum_{t=1}^T r (\mu^{N}_t)
        \to \sum_{t=1}^T r (\mu_t) \quad \textrm{in probability for} \quad N \to \infty \, .
    \end{align*}
    The reward $r$ is a continuous function by Assumption \ref{as:cont_reward}. Furthermore, by Theorem \ref{thm:mf_conv} we know that the empirical mean fields converge in probability to the limiting mean fields. Hence, we can apply the continuous mapping theorem \citep{mann1943stochastic, van2000asymptotic} to obtain the desired result.
\end{proof}

\subsection{Proof of Corollary \ref{cor:pol_opt}}

\begin{proof}
    Quantify the gap $\Delta$ between the optimal and the second best solution as
    \begin{align*}
        \Delta \coloneqq J (\pi_1) - \max_{i \in [M], i \neq 1} J (\pi_i) > 0 \, .
    \end{align*}
    Keeping in mind Proposition \ref{prop:obj_conv}, we know that the objectives of the finite systems eventually converge to the limiting mean field objectives as $N$ approaches infinity. Thus, there exists some $N^*$ such that
    \begin{align*}
        \max_{i \in [M]} \lvert J^{N} (\pi_i) - J (\pi_i) \rvert < \frac{\Delta}{2}
    \end{align*}
    holds for all $N > N^*$. Finally, the above considerations allow us to bound the difference of interest
    \begin{align*}
        &J^{N} (\pi_1) - \max_{i \in [M], i \neq 1} J^{N} (\pi_i) \\
        &\quad = J^{N} (\pi_1) - J (\pi_1) + J (\pi_1) - \max_{i \in [M], i \neq 1} J^{N} (\pi_i) \\
        &\quad = \underbrace{J^{N} (\pi_1) - J (\pi_1)}_{> - \Delta / 2} + \underbrace{J (\pi_1) - \left( \max_{i \in [M], i \neq 1} J (\pi_i) \right)}_{= \Delta} + \left( \max_{i \in [M], i \neq 1} J (\pi_i) \right) - \max_{i \in [M], i \neq 1} J^{N} (\pi_i) \\
        &\quad > \frac{\Delta}{2} + \min_{i \in [M], i \neq 1} J (\pi_i) - J^{N} (\pi_i) > \frac{\Delta}{2} - \frac{\Delta}{2} = 0,
    \end{align*}
    for all $N > N^*$ which implies the desired statement
    \begin{align*}
        J^{N} (\pi_1) > \max_{i \in [M], i \neq 1} J^{N} (\pi_i)
    \end{align*}
    and thereby concludes the proof.
\end{proof}

\subsection{Proof of Lemma \ref{lem:num_neigh}}

\begin{proof}
    Since we want to lower bound the number of possible $t$-hop neighborhoods $N_{G,t}$, we assume for simplicity that $t$-hop neighbors of the initial agent have at most degree $k$ themselves. Furthermore, we keep in mind the fact \citep[Theorem 2.2]{beck2007computing} that in a $d$-dimensional simplex with edge length $\ell \in \mathbb{N}$, the number of integer points contained in the simplex is
    \begin{align}
        \binom{d + \ell}{d} = \frac{(d + \ell)!}{d! \ell!} \, .
        \label{eq:pascal_simplex}
    \end{align}
    Since Lemma \ref{lem:num_neigh} considers the worst case, it suffices to prove the lower bound $\Omega \left( 2^{\mathrm{poly} (k)} \right)$ for one class of CL graphs. We choose our running example of CL power law graphs with coefficient above two. It is well known that these large power law CL graphs are locally tree-like \citep[Theorem 3.18]{van2024random} which we tacitly exploit in the following induction proof.
    
    The proof is via induction over $t$. We start with $t =1$ and the corresponding $1$-hop neighborhood. The neighborhood consists of $k$ agents where each one has a degree in $[k]$ and a state in $\mathcal{X}$. Since the agents themselves are indistinguishable in our model, we focus on the degree-state neighborhood distributions. Then, the set of possible state-degree neighborhood distributions can be seen as the integer points in a $(k + \vert \mathcal{X} \vert -1)$-dimensional simplex with edge length $k$.
    Keeping in mind Equation \eqref{eq:pascal_simplex} and the well-known Stirling approximation, see e.g. \citet{marsaglia1990new}, we obtain
    \begin{align*}
        N_{G,t} &\geq \binom{k + \vert \mathcal{X} \vert -1 + k}{k + \vert \mathcal{X} \vert -1} \\
        &= \frac{(2k + \vert \mathcal{X} \vert -1)!}{(k + \vert \mathcal{X} \vert -1)! k!} \\
        &\overset{\mathrm{Stirling}}{\sim} \sqrt{\frac{2 \pi (2k + \vert \mathcal{X} \vert -1)}{2 \pi (k + \vert \mathcal{X} \vert -1) 2 \pi k}} \frac{(2k + \vert \mathcal{X} \vert -1)^{2k + \vert \mathcal{X} \vert -1}}{(k + \vert \mathcal{X} \vert -1)^{k + \vert \mathcal{X} \vert -1} k^k} \\
        &\geq \frac{1}{\sqrt{2 \pi k}} \frac{(2k + \vert \mathcal{X} \vert -1)^{k}}{k^k} \\
        &\geq \frac{2^k}{\sqrt{2 \pi k}} \\
        &= 2^{k - 1/2 \cdot \log_2 (2 \pi k)} \in \Theta \left(2^{\mathrm{poly} (k)} \right).
    \end{align*}
    Now, it remains to establish the induction step from $t$ to $t+1$ where we assume that $N_{G,t} \in \Omega \left( 2^{\mathrm{poly} (k)} \right)$ holds. Then, instead of looking at the $(t+1)$-hop neighborhoods of the initial agent, we can equivalently look at his or her $1$-hop neighborhoods where each neighbor's 'extended state' now consists of the neighbor's $t$-hop neighborhood, where we ignore the edge between the neighbor and initial agent. Thus, the simplex edge length decreases by one from $k$ to $k-1$ which is negligible for large $k$. Leveraging the induction assumption, we obtain
    \begin{align*}
        N_{G, t+1} &= \Omega \left( \binom{N_{G,t} + k}{k}\right) \\
        &= \Omega \left( \frac{(N_{G,t} + k)!}{k! N_{G,t}!}\right) \\
        &\overset{\mathrm{Stirling}}{=} \Omega \left( \sqrt{\frac{N_{G,t} + k}{k N_{G,t}}} \frac{(N_{G,t} + k)^{N_{G,t} + k}}{N_{G,t}^{N_{G,t}} k^k} \right) \\
        &= \Omega \left( \frac{1}{\sqrt{k}} \frac{(N_{G,t} + k)^{k}}{k^k} \right) \\
        &\overset{\mathrm{(IA)}}{=} \Omega \left( \frac{1}{\sqrt{k}} \frac{\left( 2^{\mathrm{poly} (k)} \right)^{k}}{k^k} \right) \\
        &= \Omega \left( 2^{k \cdot \mathrm{poly} (k) - (k+1/2) \log_2 (k)} \right) \\
        &= \Omega \left( 2^{\mathrm{poly} (k)} \right)
    \end{align*}
    which concludes the proof.
\end{proof}

\section{Extensive Approximation Derivation} \label{app:ext_approx_deriv}

The goal of the following section is to establish a detailed approximation of the probability
\begin{align*}
    P_{\boldsymbol{\pi}, \boldsymbol{\mu}} \left( \mathbb{G}_{t + 1}^{k}  \left( {\boldsymbol \mu}_{t} \right) = G, x_{t+1} = x \right) \, .
\end{align*}
First, we condition on the previous neighborhood distribution and the previous state of the agent at time $t$
\begin{align*}
    &P_{\boldsymbol{\pi}, \boldsymbol{\mu}} \left( \mathbb{G}_{t + 1}^{k}  \left( {\boldsymbol \mu}_{t} \right) = G, x_{t+1} = x \right) \\
    &\qquad = \sum_{x' \in \mathcal X} P_{\boldsymbol{\pi}, \boldsymbol{\mu}} \left( \mathbb{G}_{t + 1}^{k}  \left( {\boldsymbol \mu}_{t} \right) = G, x_{t+1} = x, x_t = x' \right) \\
    &\qquad = \sum_{G' \in \boldsymbol{\mathcal{G}}^k} \sum_{x' \in \mathcal X} P_{\boldsymbol{\pi}, \boldsymbol{\mu}} \left( \mathbb{G}_{t + 1}^{k}  \left( {\boldsymbol \mu}_{t} \right) = G, \mathbb{G}_{t}^{k}  \left( {\boldsymbol \mu}_{t} \right) = G', x_{t+1} = x, x_t = x' \right) \, .
\end{align*}
Now, we can decompose the above expression into three separate terms
\begin{align*}
    & P_{\boldsymbol{\pi}, \boldsymbol{\mu}} \left( \mathbb{G}_{t + 1}^{k}  \left( {\boldsymbol \mu}_{t} \right) = G, \mathbb{G}_{t}^{k}  \left( {\boldsymbol \mu}_{t} \right) = G', x_{t+1} = x, x_t = x' \right) \\
    &\qquad = \underbrace{P_{\boldsymbol{\pi}, \boldsymbol{\mu}} \left( \mathbb{G}_{t}^{k}  \left( {\boldsymbol \mu}_{t} \right) = G', x_t = x' \right)}_{\mathrm{(I)}} \cdot
    \underbrace{P_{\boldsymbol{\pi}, \boldsymbol{\mu}} \left( x_{t+1} = x \mid \mathbb{G}_{t}^{k}  \left( {\boldsymbol \mu}_{t} \right) = G', x_t = x' \right)}_{\mathrm{(II)}} \\
    &\qquad \qquad\qquad\qquad\qquad\qquad\qquad \cdot \underbrace{P_{\boldsymbol{\pi}, \boldsymbol{\mu}} \left( \mathbb{G}_{t + 1}^{k}  \left( {\boldsymbol \mu}_{t} \right) = G \mid \mathbb{G}_{t}^{k}  \left( {\boldsymbol \mu}_{t} \right) = G', x_{t+1} = x, x_t = x' \right)}_{\mathrm{(III)}}
\end{align*}
which allows us to handle each term individually. Since we require a recursive computation of the probability $P_{\boldsymbol{\pi}, \boldsymbol{\mu}} \left( \mathbb{G}_{t + 1}^{k}  \left( {\boldsymbol \mu}_{t} \right) = G, x_{t+1} = x \right)$, the first term (I) will not be reformulated any further. The computation of the second term (II) is straight-forward, i.e.
\begin{align*}
    P_{\boldsymbol{\pi}, \boldsymbol{\mu}} \left( x_{t+1} = x \mid \mathbb{G}_{t}^{k}  \left( {\boldsymbol \mu}_{t} \right) = G', x_t = x' \right) = \sum_{u \in \mathcal{U}} \pi^k \left( u \mid x' \right) P \left( x \mid x', u, G'\right) \, .
\end{align*}
Thus, it remains to approximate the third term (III)
\begin{align*}
    &P_{\boldsymbol{\pi}, \boldsymbol{\mu}} \left( \mathbb{G}_{t + 1}^{k}  \left( {\boldsymbol \mu}_{t} \right) = G \mid \mathbb{G}_{t}^{k}  \left( {\boldsymbol \mu}_{t} \right) = G', x_{t+1} = x, x_t = x' \right) \\
    &\qquad \qquad \qquad \qquad \qquad \qquad \qquad \qquad \qquad =P_{\boldsymbol{\pi}, \boldsymbol{\mu}} \left( \mathbb{G}_{t + 1}^{k}  \left( {\boldsymbol \mu}_{t} \right) = G \mid \mathbb{G}_{t}^{k}  \left( {\boldsymbol \mu}_{t} \right) = G', x_t = x' \right) \, .
\end{align*}
To ensure a reasonable approximation complexity, we make the simplifying assumption that the neighborhood distribution does not (crucially) depend on the current state of the agent of interest, i.e.
\begin{align*}
    P_{\boldsymbol{\pi}, \boldsymbol{\mu}} \left( \mathbb{G}_{t + 1}^{k}  \left( {\boldsymbol \mu}_{t} \right) = G \mid \mathbb{G}_{t}^{k}  \left( {\boldsymbol \mu}_{t} \right) = G', x_t = x' \right) \approx 
    P_{\boldsymbol{\pi}, \boldsymbol{\mu}} \left( \mathbb{G}_{t + 1}^{k}  \left( {\boldsymbol \mu}_{t} \right) = G \mid \mathbb{G}_{t}^{k}  \left( {\boldsymbol \mu}_{t} \right) = G' \right) \, .
\end{align*}
Thus, we focus on
\begin{align*}
    P_{\boldsymbol{\pi}, \boldsymbol{\mu}} \left( \mathbb{G}_{t + 1}^{k}  \left( {\boldsymbol \mu}_{t} \right) = G \mid \mathbb{G}_{t}^{k}  \left( {\boldsymbol \mu}_{t} \right) = G' \right)
\end{align*}
which requires an involved combinatorial argument to be calculated. The main difficulty in the calculation stems from the fact that the $k$ neighbors of the initial agent in general have different degrees, different states at time $t$ as well as different states at time $t+1$.
For notational convenience, we denote by $x_{1, t}, \ldots, x_{k,t}$ the states of the $k$ neighbors of the initial agent at time $t$ and by $\deg_1, \ldots, \deg_{k^*}, \deg_{\infty} \in \{1, \ldots, k \}$ the number of neighbors with the respective degree. Also, define $\boldsymbol{\mathcal{C}}^k \coloneqq \{ c= (c_1, \ldots, c_{k^*}, c_{\infty}) \in \mathbb{N}_0^{k^* +1}: c_1 + \ldots + c_{k^*} + c_{\infty} = k \}$ for notational convenience. Then, the above probability can be expressed as
\begin{align*}
    &P_{\boldsymbol{\pi}, \boldsymbol{\mu}} \left( \mathbb{G}_{t + 1}^{k}  \left( {\boldsymbol \mu}_{t} \right) = G \mid \mathbb{G}_{t}^{k}  \left( {\boldsymbol \mu}_{t} \right) = G' \right) \\
    &= \sum_{c \in \boldsymbol{\mathcal{C}}^k} P_{\boldsymbol{\pi}, \boldsymbol{\mu}} \left( \mathbb{G}_{t + 1}^{k} \left( {\boldsymbol \mu}_{t} \right) = G, \deg_1 = c_1, \ldots, \deg_{k^*} = c_{k^*}, \deg_{\infty} = c_{\infty} \mid \mathbb{G}_{t}^{k} \left( {\boldsymbol \mu}_{t} \right) = G' \right) \\
    &= \sum_{c \in \boldsymbol{\mathcal{C}}^k} P_{\boldsymbol{\pi}, \boldsymbol{\mu}} \left( \deg_1 = c_1, \ldots, \deg_{k^*} = c_{k^*}, \deg_{\infty} = c_{\infty} \mid \mathbb{G}_{t}^{k}  \left( {\boldsymbol \mu}_{t} \right) = G' \right) \\
    &\qquad \qquad \cdot P_{\boldsymbol{\pi}, \boldsymbol{\mu}} \left( \mathbb{G}_{t + 1}^{k}  \left( {\boldsymbol \mu}_{t} \right) = G \mid \deg_1 = c_1, \ldots, \deg_{k^*} = c_{k^*}, \deg_{\infty} = c_{\infty}, \mathbb{G}_{t}^{k} \left( {\boldsymbol \mu}_{t} \right) = G' \right) \, .
\end{align*}

In the remainder of the derivation, we will frequently use for all $s \in \mathcal{X}, m \in \mathbb{N}$, and $t \in \mathcal{T}$ the approximation
\begin{align}\label{heur_ap:state_deg}
    P \left(x_t^1 = s \mid \deg (v_1) = m, (v_0, v_1) \in E \right) \approx P \left(x_t^1 = s \mid \deg (v_1) = m \right) = \mu_t^m (s) \, .
\end{align}
Next, we make an auxiliary calculation to calculate the degree distribution of a (uniformly at random picked) node $v_1$ conditional on its state $x_t^1$ and that it is a neighbor of the initial node $v_0$ of interest
\begin{align*}
    &P \left(\deg (v_1) = m \mid x_t^1 = s, (v_0, v_1) \in E \right) \\
    &= \frac{P \left(\deg (v_1) = m \cap x_t^1 = s \mid (v_0, v_1) \in E \right)}{P \left(x_t^1 = s \mid (v_0, v_1) \in E \right)} \\
    &= \frac{P \left(\deg (v_1) = m \mid (v_0, v_1) \in E \right) P \left(x_t^1 = s \mid \deg (v_1) = m, (v_0, v_1) \in E \right)}{P \left(\deg (v_1) > k^* \cap x_t^1 = s \mid (v_0, v_1) \in E \right) + \sum_{k=1}^{k^*} P \left(\deg (v_1) = k \cap x_t^1 = s \mid (v_0, v_1) \in E \right)} \\
    &\substack{\eqref{heur_ap:state_deg} \\ \approx} \frac{P \left(\deg (v_1) = m \mid (v_0, v_1) \in E \right) \mu_t^m (s)}{P \left(\deg (v_1) > k^* \cap x_t^1 = s \mid (v_0, v_1) \in E \right) + \sum_{k=1}^{k^*} P \left(\deg (v_1) = k \cap x_t^1 = s \mid (v_0, v_1) \in E \right)} \\
    &= \frac{P \left(\deg (v_1) = m \mid (v_0, v_1) \in E \right) \mu_t^m (s)}{P \left(\deg (v_1) > k^*\mid (v_0, v_1) \in E \right) \mu_t^\infty (s) + \sum_{k=1}^{k^*} P \left(\deg (v_1) = k \mid (v_0, v_1) \in E \right) \mu_t^k (s)}
\end{align*}
where we exploit that
\begin{align*}
    &P \left(\deg (v_1) > k^* \cap x_t^1 = s \mid (v_0, v_1) \in E \right) + \sum_{k=1}^{k^*} P \left(\deg (v_1) = k \cap x_t^1 = s \mid (v_0, v_1) \in E \right) \\
    &\qquad = P \left(\deg (v_1) > k^*\mid (v_0, v_1) \in E \right) P \left(x_t^1 = s \mid \deg (v_1) > k^*, (v_0, v_1) \in E \right) \\
    &\qquad \qquad + \sum_{k=1}^{k^*} P \left(\deg (v_1) = k \mid (v_0, v_1) \in E \right) P \left(x_t^1 = s \mid \deg (v_1) = k, (v_0, v_1) \in E \right) \\
    &\qquad \substack{\eqref{heur_ap:state_deg} \\ \approx} P \left(\deg (v_1) > k^*\mid (v_0, v_1) \in E \right) \mu_t^\infty (s) + \sum_{k=1}^{k^*} P \left(\deg (v_1) = k \mid (v_0, v_1) \in E \right) \mu_t^k (s) \, .
\end{align*}
For the running example of power law degree distributions with exponent $\gamma \in (2,3)$, the conditional degree distribution is approximately
\begin{align*}
    &P \left(\deg (v_1) = m \mid x_t^1 = s_j, (v_0, v_1) \in E \right) \\
    &\qquad \qquad \qquad \qquad \qquad \approx \frac{ \frac{m^{1 - \gamma}}{\zeta (\gamma - 1)} \mu_t^m (s_j)}{\frac{1}{\zeta (\gamma - 1)}\left[ \sum_{\ell = k^* + 1}^{\infty} \ell^{1 - \gamma} \right] \mu_t^\infty (s_j) + \frac{1}{\zeta (\gamma - 1)} \sum_{h=1}^{k^*} h^{1 - \gamma} \mu_t^h (s_j)} \\
    &\qquad \qquad \qquad \qquad \qquad = \frac{m^{1 - \gamma} \mu_t^m (s_j)}{\left[ \sum_{\ell = k^* + 1}^{\infty} \ell^{1 - \gamma} \right] \mu_t^\infty (s_j) + \sum_{h=1}^{k^*} h^{1 - \gamma} \mu_t^h (s_j)} \, .
\end{align*}
Based on the above probability and by the symmetry of the model, we obtain
\begin{align*}
    &P_{\boldsymbol{\pi}, \boldsymbol{\mu}} \left( \deg_1 = c_1, \ldots, \deg_{k^*} = c_{k^*}, \deg_{\infty} = c_{\infty} \mid \mathbb{G}_{t}^{k}  \left( {\boldsymbol \mu}_{t} \right) = G' \right) \\
    &\quad = \sum_{\boldsymbol{a}_2 \in \boldsymbol{\mathcal{A}}^k_2 (G', c)} P_{\boldsymbol{\pi}, \boldsymbol{\mu}} \left( A_2 = \boldsymbol{a}_2, \deg_1 = c_1, \ldots, \deg_{k^*} = c_{k^*}, \deg_{\infty} = c_{\infty} \mid \mathbb{G}_{t}^{k}  \left( {\boldsymbol \mu}_{t} \right) = G' \right) \\
    &\quad = \sum_{\boldsymbol{a}_2 \in \boldsymbol{\mathcal{A}}^k_2 (G', c)} P_{\boldsymbol{\pi}, \boldsymbol{\mu}} \left( A_2 = \boldsymbol{a}_2 \mid \mathbb{G}_{t}^{k}  \left( {\boldsymbol \mu}_{t} \right) = G' \right) \\
    &\quad \approx \sum_{\boldsymbol{a}_2 \in \boldsymbol{\mathcal{A}}^k_2 (G', c)} \prod_{j=1}^{d} \binom{g_j'}{a_{j 1}, \ldots, a_{j \infty}} \prod_{m \in [k^*] \cup \{ \infty \}}  ( P \left(\deg (v_1) = m \mid x_t^1 = s_j, (v_0, v_1) \in E \right) )^{a_{j m}}\\
\end{align*}
where we neglect dependencies between the nodes in the last line and define the matrix set $\boldsymbol{\mathcal{A}}^k_2 (G', c)$ for given $G' \in \boldsymbol{\mathcal{G}}^k$ and $c \in \boldsymbol{\mathcal{C}}^k$ as 
\begin{align*}
    &\boldsymbol{\mathcal{A}}^k_2 (G', c) \coloneqq \left\{ \boldsymbol{a}_2 = (a_{jm})_{j \in [d], m \in [k^*] \cup \{ \infty \}} \in \mathbb{N}_0^{d \times (k^* +1)}: \right. \\
    &\qquad \qquad \qquad \left. \sum_{m' \in [k^*] \cup \{ \infty \}} a_{j m'} = g'_j, \forall j \in [d] \quad \textrm{and} \quad \sum_{\ell=1}^d a_{\ell  m} = c_m, \forall m \in [k^*] \cup \{ \infty \} \right\} \, .
\end{align*}
Therefore, it remains to calculate the conditional probability
\begin{align*}
    P_{\boldsymbol{\pi}, \boldsymbol{\mu}} \left( \mathbb{G}_{t + 1}^{k}  \left( {\boldsymbol \mu}_{t} \right) = G \mid \deg_1 = c_1, \ldots, \deg_{k^*} = c_{k^*}, \deg_{\infty} = c_{\infty}, \mathbb{G}_{t}^{k} \left( {\boldsymbol \mu}_{t} \right) = G' \right) \, .
\end{align*}
As a first step, we define the set of matrices $\boldsymbol{\mathcal{A}}^k_3 (G, G', c)$ for a given triple of vectors $G, G' \in \boldsymbol{\mathcal{G}}^k$ and $c \in \boldsymbol{\mathcal{C}}^k$ as 
\begin{align*}
    &\boldsymbol{\mathcal{A}}^k_3 (G, G', c) \coloneqq \left\{ \boldsymbol{a}_3 = (a_{ijm})_{i, j \in [d], m \in [k^*] \cup \{ \infty \}} \in \mathbb{N}_0^{d \times d \times (k^* +1)}: \right. \\
    &\qquad \qquad \qquad \qquad  \sum_{m' \in [k^*] \cup \{ \infty \}} \sum_{\ell=1}^d a_{i \ell m'} = g_i \textrm{ and } \sum_{m' \in [k^*] \cup \{ \infty \}} \sum_{\ell=1}^d a_{\ell j m'} = g'_j, \quad \forall i,j \in [d]  \\
    & \left. \qquad \qquad \qquad \qquad \qquad \qquad \qquad \qquad \textrm{ and } \sum_{\ell, \ell' =1}^d a_{\ell \ell' m} = c_m, \forall m \in [k^*] \cup \{ \infty \} \right\} \, .
\end{align*}
where $d \coloneqq \vert \mathcal{X} \vert$ is the finite number of states. Intuitively, the matrix set $\boldsymbol{\mathcal{A}}^k_3 (G, G', c)$ for an agent with degree $k$ contains all possible numbers $(a_{ijm})_{i, j \in [d], m \in [k^*] \cup \{ \infty \}}$ of neighbors  whose degree is $m$ and current state is $x_i$
and who transition to state $x_j$ in the next time step. For notational convenience, let $A$ denote the random variable taking values in $\boldsymbol{\mathcal{A}}^k_3 (G, G', c)$ and analogously let $A_2$ be the random variable with values in $\boldsymbol{\mathcal{A}}^k_2 (G', c)$. We continue with the reformulation
\begin{align*}
    &P_{\boldsymbol{\pi}, \boldsymbol{\mu}} \left( \mathbb{G}_{t + 1}^{k}  \left( {\boldsymbol \mu}_{t} \right) = G \mid \deg_1 = c_1, \ldots, \deg_{k^*} = c_{k^*}, \deg_{\infty} = c_{\infty}, \mathbb{G}_{t}^{k} \left( {\boldsymbol \mu}_{t} \right) = G' \right) \\
    &= \sum_{\boldsymbol{a}_2 \in \boldsymbol{\mathcal{A}}^k_2 (G', c)} P_{\boldsymbol{\pi}, \boldsymbol{\mu}} \left( A_2 = \boldsymbol{a}_2 \mid \deg_1 = c_1, \ldots, \deg_{k^*} = c_{k^*}, \deg_{\infty} = c_{\infty}, \mathbb{G}_{t}^{k} \left( {\boldsymbol \mu}_{t} \right) = G' \right) \\
    &\quad \cdot P_{\boldsymbol{\pi}, \boldsymbol{\mu}} \left( \mathbb{G}_{t + 1}^{k}  \left( {\boldsymbol \mu}_{t} \right) = G \mid A_2 = \boldsymbol{a}_2, \deg_1 = c_1, \ldots, \deg_{k^*} = c_{k^*}, \deg_{\infty} = c_{\infty}, \mathbb{G}_{t}^{k} \left( {\boldsymbol \mu}_{t} \right) = G' \right) \\
    &= \sum_{\boldsymbol{a}_2 \in \boldsymbol{\mathcal{A}}^k_2 (G', c)} P_{\boldsymbol{\pi}, \boldsymbol{\mu}} \left( A_2 = \boldsymbol{a}_2 \mid \deg_1 = c_1, \ldots, \deg_{k^*} = c_{k^*}, \deg_{\infty} = c_{\infty}, \mathbb{G}_{t}^{k} \left( {\boldsymbol \mu}_{t} \right) = G' \right) \\
    &\quad \cdot P_{\boldsymbol{\pi}, \boldsymbol{\mu}} \left( \mathbb{G}_{t + 1}^{k}  \left( {\boldsymbol \mu}_{t} \right) = G \mid A_2 = \boldsymbol{a}_2 \right) \, .
\end{align*}
Next, we consider the two conditional probabilities separately. We start with
\begin{align*}
    &P_{\boldsymbol{\pi}, \boldsymbol{\mu}} \left( A_2 = \boldsymbol{a}_2 \mid \deg_1 = c_1, \ldots, \deg_{k^*} = c_{k^*}, \deg_{\infty} = c_{\infty}, \mathbb{G}_{t}^{k} \left( {\boldsymbol \mu}_{t} \right) = G' \right) \\
    &\qquad \qquad = \frac{P_{\boldsymbol{\pi}, \boldsymbol{\mu}} \left( A_2 = \boldsymbol{a}_2 \cap \deg_1 = c_1, \ldots, \deg_{k^*} = c_{k^*}, \deg_{\infty} = c_{\infty}, \mathbb{G}_{t}^{k} \left( {\boldsymbol \mu}_{t} \right) = G' \right)}{P_{\boldsymbol{\pi}, \boldsymbol{\mu}} \left( \deg_1 = c_1, \ldots, \deg_{k^*} = c_{k^*}, \deg_{\infty} = c_{\infty}, \mathbb{G}_{t}^{k} \left( {\boldsymbol \mu}_{t} \right) = G' \right)} \\
    &\qquad \qquad = \frac{P_{\boldsymbol{\pi}, \boldsymbol{\mu}} \left( A_2 = \boldsymbol{a}_2 \right)}{P_{\boldsymbol{\pi}, \boldsymbol{\mu}} \left( \deg_1 = c_1, \ldots, \deg_{k^*} = c_{k^*}, \deg_{\infty} = c_{\infty}, \mathbb{G}_{t}^{k} \left( {\boldsymbol \mu}_{t} \right) = G' \right)} \, .
\end{align*}
Keeping in mind both
\begin{align*}
    P_{\boldsymbol{\pi}, \boldsymbol{\mu}} \left( A_2 = \boldsymbol{a}_2 = (a_{j m})_{j,m} \right)
    &\approx \prod_{j=1}^{d} \prod_{m \in [k^*] \cup \{ \infty \}} \left( P \left(\deg (v_1) = m \mid (v_0, v_1) \in E \right) \mu_t^m (s_j) \right)^{a_{jm}}
\end{align*}
by neglecting dependencies between the nodes and
\begin{align*}
    &P_{\boldsymbol{\pi}, \boldsymbol{\mu}} \left( \deg_1 = c_1, \ldots, \deg_{k^*} = c_{k^*}, \deg_{\infty} = c_{\infty}, \mathbb{G}_{t}^{k} \left( {\boldsymbol \mu}_{t} \right) = G' \right) \\
    &\quad = \sum_{\boldsymbol{a}_2 \in \boldsymbol{\mathcal{A}}^k_2 (G', c)} P_{\boldsymbol{\pi}, \boldsymbol{\mu}} \left( \deg_1 = c_1, \ldots, \deg_{k^*} = c_{k^*}, \deg_{\infty} = c_{\infty}, \mathbb{G}_{t}^{k} \left( {\boldsymbol \mu}_{t} \right) = G', A_2 = \boldsymbol{a}_2 \right) \\
    &\quad = \sum_{\boldsymbol{a}_2 \in \boldsymbol{\mathcal{A}}^k_2 (G', c)} P_{\boldsymbol{\pi}, \boldsymbol{\mu}} \left( A_2 = \boldsymbol{a}_2 \right) \\
    &\quad = \sum_{\boldsymbol{a}_2 \in \boldsymbol{\mathcal{A}}^k_2 (G', c)} \prod_{j=1}^{d} \prod_{m \in [k^*] \cup \{ \infty \}} \left( P \left(\deg (v_1) = m \mid (v_0, v_1) \in E \right) \mu_t^m (s_j)\right)^{a_{jm}}
\end{align*}
we obtain
\begin{align*}
    &P_{\boldsymbol{\pi}, \boldsymbol{\mu}} \left( A_2 = \boldsymbol{a}_2 \mid \deg_1 = c_1, \ldots, \deg_{k^*} = c_{k^*}, \deg_{\infty} = c_{\infty}, \mathbb{G}_{t}^{k} \left( {\boldsymbol \mu}_{t} \right) = G' \right) \\
    &\qquad = \frac{P_{\boldsymbol{\pi}, \boldsymbol{\mu}} \left( A_2 = \boldsymbol{a}_2 \right)}{P_{\boldsymbol{\pi}, \boldsymbol{\mu}} \left( \deg_1 = c_1, \ldots, \deg_{k^*} = c_{k^*}, \deg_{\infty} = c_{\infty}, \mathbb{G}_{t}^{k} \left( {\boldsymbol \mu}_{t} \right) = G' \right)} \\
    &\qquad \approx \frac{\prod_{j=1}^{d} \prod_{m \in [k^*] \cup \{ \infty \}} \left( P \left(\deg (v_1) = m \mid (v_0, v_1) \in E \right) \mu_t^m (s_j) \right)^{a_{jm}}}{\sum_{ \boldsymbol{a}_2' \in \boldsymbol{\mathcal{A}}^k_2 (G', c)} \prod_{j=1}^{d} \prod_{m \in [k^*] \cup \{ \infty \}} \left( P \left(\deg (v_1) = m \mid (v_0, v_1) \in E \right) \mu_t^m (s_j)\right)^{a'_{jm}}}
\end{align*}
and especially, for the case of a power law degree distribution with $\gamma \in (2,3)$, we have
\begin{align*}
    &P_{\boldsymbol{\pi}, \boldsymbol{\mu}} \left( A_2 = \boldsymbol{a}_2 \mid \deg_1 = c_1, \ldots, \deg_{k^*} = c_{k^*}, \deg_{\infty} = c_{\infty}, \mathbb{G}_{t}^{k} \left( {\boldsymbol \mu}_{t} \right) = G' \right) \\
    &= \frac{\prod_{j=1}^{d} \left( \mu_t^\infty (s_j)  \left( 1- \sum_{m' =1}^{k^*} \frac{(m')^{1- \gamma}}{\zeta (\gamma - 1)} \right) \right)^{a_{j \infty}}
    \prod_{m=1}^{k^*} \left( \frac{m^{1- \gamma} \mu_t^m (s_j)}{\zeta (\gamma - 1)}\right)^{a_{jm}}}{\sum_{\boldsymbol{a}_2' \in \boldsymbol{\mathcal{A}}^k_2 (G', c)} \prod_{j=1}^{d} \left( \mu_t^\infty (s_j)  \left( 1- \sum_{m' =1}^{k^*} \frac{(m')^{1- \gamma}}{\zeta (\gamma - 1)} \right) \right)^{a'_{j \infty}} \prod_{m=1}^{k^*} \left( \frac{m^{1- \gamma} \mu_t^m (s_j)}{\zeta (\gamma - 1)}\right)^{a'_{jm}}} \\
    &\approx \frac{\prod_{j=1}^{d} \left( \mu_t^\infty (s_j)  \left( \zeta (\gamma - 1) - \sum_{m' =1}^{k^*} (m')^{1- \gamma} \right) \right)^{a_{j \infty}}
    \prod_{m=1}^{k^*} \left( m^{1- \gamma} \mu_t^m (s_j) \right)^{a_{jm}}}{\sum_{\boldsymbol{a}_2' \in \boldsymbol{\mathcal{A}}^k_2 (G', c)} \prod_{j=1}^{d} \left( \mu_t^\infty (s_j)  \left( \zeta (\gamma - 1) - \sum_{m' =1}^{k^*} (m')^{1- \gamma} \right) \right)^{a'_{j \infty}} \prod_{m=1}^{k^*} \left( m^{1- \gamma} \mu_t^m (s_j) \right)^{a'_{jm}}} .
\end{align*}
Now, it remains to calculate the second probability term, namely 
\begin{align*}
    P_{\boldsymbol{\pi}, \boldsymbol{\mu}} \left( \mathbb{G}_{t + 1}^{k}  \left( {\boldsymbol \mu}_{t} \right) = G \mid A_2 = \boldsymbol{a}_2 \right) \, .
\end{align*}
Exploiting the symmetry of the problem, we obtain
\begin{align*}
    &P_{\boldsymbol{\pi}, \boldsymbol{\mu}} \left( \mathbb{G}_{t + 1}^{k}  \left( {\boldsymbol \mu}_{t} \right) = G \mid A_2 = \boldsymbol{a}_2 \right) \\
    &\quad \approx \sum_{\boldsymbol{a}_3 \in \boldsymbol{\mathcal{A}}^k (G, G', c)} \prod_{j=1}^{d} \prod_{m \in [k^*] \cup \{ \infty \}} \binom{\sum_i a_{i j m}}{a_{1 j m}, \ldots, a_{d j m}} \boldsymbol{1}_{\{ \sum_i a_{ijm} = a_{jm} \}}\\
    &\quad \qquad \cdot \prod_{i=1}^{d} \left( P_{\boldsymbol{\pi}, \boldsymbol{\mu}} \left( x_{t+1}^1 = x_i \mid x_t^1 = x_j, \deg (v_1) = m \right) \right)^{a_{i j m}} \\
    &\quad \approx \sum_{\boldsymbol{a}_3 \in \boldsymbol{\mathcal{A}}^k (G, G', c)} \prod_{j=1}^{d} \prod_{m \in [k^*] \cup \{ \infty \}} \binom{\sum_i a_{i j m}}{a_{1 j m}, \ldots, a_{d j m}} \boldsymbol{1}_{\{ \sum_i a_{ijm} = a_{jm} \}}\\
    &\quad \qquad \cdot \prod_{i=1}^{d} \left( \sum_{G'' \in \boldsymbol{\mathcal{G}}^m} P_{\boldsymbol{\pi}} \left( \mathbb{G}_{t}^{m} \left( {\boldsymbol \mu}_{t} \right) = G'' \mid x''_t = s_j \right)  \sum_{u \in \mathcal{U}} \pi_{t}^m \left(u \mid s_{j} \right) \cdot P \left( s_i \mid s_{j}, u, G'' \right) \right)^{a_{i j m}}
\end{align*}
where $\boldsymbol{1}_{\{ \ldots \}}$ denotes the indicator function and where we neglect the potential dependencies between the neighbors of the initial node in the second line. Finally, we arrive at
\begin{align*}
    &P_{\boldsymbol{\pi}, \boldsymbol{\mu}} \left( \mathbb{G}_{t + 1}^{k}  \left( {\boldsymbol \mu}_{t} \right) = G \mid \deg_1 = c_1, \ldots, \deg_{k^*} = c_{k^*}, \deg_{\infty} = c_{\infty}, \mathbb{G}_{t}^{k} \left( {\boldsymbol \mu}_{t} \right) = G' \right) \\
    &= \sum_{\boldsymbol{a}_2 \in \boldsymbol{\mathcal{A}}^k_2 (G', c)} P_{\boldsymbol{\pi}, \boldsymbol{\mu}} \left( A_2 = \boldsymbol{a}_2 \mid \deg_1 = c_1, \ldots, \deg_{k^*} = c_{k^*}, \deg_{\infty} = c_{\infty}, \mathbb{G}_{t}^{k} \left( {\boldsymbol \mu}_{t} \right) = G' \right) \\
    &\qquad \qquad \qquad \qquad \qquad \qquad \qquad \qquad \qquad \qquad \qquad \qquad \cdot P_{\boldsymbol{\pi}, \boldsymbol{\mu}} \left( \mathbb{G}_{t + 1}^{k}  \left( {\boldsymbol \mu}_{t} \right) = G \mid A_2 = \boldsymbol{a}_2 \right) \\
    &\approx \sum_{\boldsymbol{a}_2 \in \boldsymbol{\mathcal{A}}^k_2 (G', c)} \frac{\prod_{j=1}^{d} \prod_{m \in [k^*] \cup \{ \infty \}} \left( P \left(\deg (v_1) = m \mid (v_0, v_1) \in E \right) \mu_t^m (s_j) \right)^{a_{jm}}}{\sum_{\boldsymbol{a}_2' \in \boldsymbol{\mathcal{A}}^k_2 (G', c)} \prod_{j=1}^{d} \prod_{m \in [k^*] \cup \{ \infty \}} \left( P \left(\deg (v_1) = m \mid (v_0, v_1) \in E \right) \mu_t^m (s_j)\right)^{a_{jm}'}}\\
    &\qquad \qquad \qquad \qquad \qquad \qquad \qquad \qquad \qquad \qquad \qquad \qquad \cdot P_{\boldsymbol{\pi}, \boldsymbol{\mu}} \left( \mathbb{G}_{t + 1}^{k}  \left( {\boldsymbol \mu}_{t} \right) = G \mid A_2 = \boldsymbol{a}_2 \right) \\
    &\approx  \sum_{\boldsymbol{a}_2 \in \boldsymbol{\mathcal{A}}^k_2 (G', c)} \frac{\prod_{j=1}^{d} \prod_{m \in [k^*] \cup \{ \infty \}} \left( P \left(\deg (v_1) = m \mid (v_0, v_1) \in E \right) \mu_t^m (s_j) \right)^{a_{jm}}}{\sum_{\boldsymbol{a}_2' \in \boldsymbol{\mathcal{A}}^k_2 (G', c)} \prod_{j=1}^{d} \prod_{m \in [k^*] \cup \{ \infty \}} \left( P \left(\deg (v_1) = m \mid (v_0, v_1) \in E \right) \mu_t^m (s_j)\right)^{a_{jm}'}} \\
    &\qquad \sum_{\boldsymbol{a}_3 \in \boldsymbol{\mathcal{A}}^k_3 (G, G', c)} \prod_{j=1}^{d} \prod_{m \in [k^*] \cup \{ \infty \}} \binom{\sum_i a_{i j m}}{a_{1 j m}, \ldots, a_{d j m}} \boldsymbol{1}_{\{ \sum_i a_{ijm} = a_{jm} \}}\\
    &\quad \qquad \cdot \prod_{i=1}^{d} \left( \sum_{G'' \in \boldsymbol{\mathcal{G}}^m} P_{\boldsymbol{\pi}} \left( \mathbb{G}_{t}^{m} \left( {\boldsymbol \mu}_{t} \right) = G'' \mid x''_t = s_j \right)  \sum_{u \in \mathcal{U}} \pi_{t}^m \left(u \mid s_{j} \right) \cdot P \left( s_i \mid s_{j}, u, G'' \right) \right)^{a_{i j m}}
\end{align*}
and for the running example of power law graphs we especially obtain
\begin{align*}
    &P_{\boldsymbol{\pi}, \boldsymbol{\mu}} \left( \mathbb{G}_{t + 1}^{k}  \left( {\boldsymbol \mu}_{t} \right) = G \mid \deg_1 = c_1, \ldots, \deg_{k^*} = c_{k^*}, \deg_{\infty} = c_{\infty}, \mathbb{G}_{t}^{k} \left( {\boldsymbol \mu}_{t} \right) = G' \right) \\
    &\approx \sum_{\boldsymbol{a}_2 \in \boldsymbol{\mathcal{A}}^k_2 (G', c)} \frac{\prod_{j=1}^{d} \left( \mu_t^\infty (s_j)  \left( 1- \sum_{m' =1}^{k^*} \frac{(m')^{1- \gamma}}{\zeta (\gamma - 1)} \right) \right)^{a_{j \infty}}
    \prod_{m=1}^{k^*} \left( \frac{m^{1- \gamma} \mu_t^m (s_j)}{\zeta (\gamma - 1)}\right)^{a_{jm}}}{\sum_{\boldsymbol{a}_2' \in \boldsymbol{\mathcal{A}}^k_2 (G', c)} \prod_{j} \left( \mu_t^\infty (s_j)  \left( 1- \sum_{m' =1}^{k^*} \frac{(m')^{1- \gamma}}{\zeta (\gamma - 1)} \right) \right)^{a'_{j \infty}} \prod_{m=1}^{k^*} \left( \frac{m^{1- \gamma} \mu_t^m (s_j)}{\zeta (\gamma - 1)}\right)^{a'_{jm}}}\\
    &\qquad \sum_{\boldsymbol{a}_3 \in \boldsymbol{\mathcal{A}}^k_3 (G, G', c)} \prod_{j=1}^{d} \prod_{m \in [k^*] \cup \{ \infty \}} \binom{\sum_i a_{i j m}}{a_{1 j m}, \ldots, a_{d j m}} \boldsymbol{1}_{\{ \sum_i a_{ijm} = a_{jm} \}}\\
    &\quad \qquad \cdot \prod_{i=1}^{d} \left( \sum_{G'' \in \boldsymbol{\mathcal{G}}^m} P_{\boldsymbol{\pi}} \left( \mathbb{G}_{t}^{m} \left( {\boldsymbol \mu}_{t} \right) = G'' \mid x''_t = s_j \right)  \sum_{u \in \mathcal{U}} \pi_{t}^m \left(u \mid s_{j} \right) \cdot P \left( s_i \mid s_{j}, u, G'' \right) \right)^{a_{i j m}} \\
    &= \sum_{\boldsymbol{a}_3 \in \boldsymbol{\mathcal{A}}^k_3 (G, G', c)} \frac{\prod_{j=1}^{d} \left( \mu_t^\infty (s_j)  \left( 1- \sum_{m' =1}^{k^*} \frac{(m')^{1- \gamma}}{\zeta (\gamma - 1)} \right) \right)^{a_{j \infty}}
    \prod_{m=1}^{k^*} \left( \frac{m^{1- \gamma} \mu_t^m (s_j)}{\zeta (\gamma - 1)}\right)^{a_{jm}}}{\sum_{\boldsymbol{a}_2' \in \boldsymbol{\mathcal{A}}^k_2 (G', c)} \prod_{j} \left( \mu_t^\infty (s_j)  \left( 1- \sum_{m' =1}^{k^*} \frac{(m')^{1- \gamma}}{\zeta (\gamma - 1)} \right) \right)^{a'_{j \infty}} \prod_{m=1}^{k^*} \left( \frac{m^{1- \gamma} \mu_t^m (s_j)}{\zeta (\gamma - 1)}\right)^{a'_{jm}}}\\
    &\qquad  \prod_{j=1}^{d} \prod_{m \in [k^*] \cup \{ \infty \}} \binom{\sum_i a_{i j m}}{a_{1 j m}, \ldots, a_{d j m}} \\
    &\quad \qquad \cdot \prod_{i=1}^{d} \left( \sum_{G'' \in \boldsymbol{\mathcal{G}}^m} P_{\boldsymbol{\pi}} \left( \mathbb{G}_{t}^{m} \left( {\boldsymbol \mu}_{t} \right) = G'' \mid x''_t = s_j \right)  \sum_{u \in \mathcal{U}} \pi_{t}^m \left(u \mid s_{j} \right) \cdot P \left( s_i \mid s_{j}, u, G'' \right) \right)^{a_{i j m}} \, .
\end{align*}

\paragraph{Resulting Approximation}
Eventually, we obtain the approximation
\begin{align*}
    &P_{\boldsymbol{\pi}, \boldsymbol{\mu}} \left( \mathbb{G}_{t + 1}^{k}  \left( {\boldsymbol \mu}_{t} \right) = G, x_{t+1} = x \right) \\
    &\quad \approx \sum_{G' \in \boldsymbol{\mathcal{G}}^k} \sum_{x' \in \mathcal X} P_{\boldsymbol{\pi}, \boldsymbol{\mu}} \left( \mathbb{G}_{t}^{k}  \left( {\boldsymbol \mu}_{t} \right) = G', x_t = x' \right) 
    \left[ \sum_{u \in \mathcal{U}} \pi^k \left( u \mid x' \right) P \left( x \mid x', u, G'\right) \right] \\
    &\quad \cdot \sum_{c \in \boldsymbol{\mathcal{C}}^k} \left[ \sum_{\boldsymbol{a}_2 \in \boldsymbol{\mathcal{A}}^k (G', c)} \prod_{j=1}^{d} \binom{g_j'}{a_{j 1}, \ldots, a_{j \infty}} \prod_{m \in [k^*] \cup \{ \infty \}} \right. \\
    &\qquad \qquad \left. \cdot \left( \frac{P \left(\deg (v_1) = m \mid (v_0, v_1) \in E \right) \mu_t^m (s_j)}{P \left(\deg (v_1) > k^*\mid (v_0, v_1) \in E \right) \mu_t^\infty (s_j) + \sum_{k=1}^{k^*} P \left(\deg (v_1) = k \mid (v_0, v_1) \in E \right) \mu_t^k (s_j)} \right)^{a_{j m}} \right] \\
    &\qquad \qquad \cdot \frac{1}{\sum_{\boldsymbol{a}_2' \in \boldsymbol{\mathcal{A}}^k_2 (G', c)} \prod_{j, m} \left( P \left(\deg (v_1) = m \mid (v_0, v_1) \in E \right) \mu_t^m (s_j) \right)^{a_{jm}'}} \\
    &\quad \qquad \sum_{\boldsymbol{a}_3 \in \boldsymbol{\mathcal{A}}^k_3 (G, G', c)} \prod_{j=1}^{d} \prod_{m \in [k^*] \cup \{ \infty \}} \binom{\sum_i a_{i j m}}{a_{1 j m}, \ldots, a_{d j m}} \prod_{i=1}^{d} \left( P \left(\deg (v_1) = m \mid (v_0, v_1) \in E \right) \mu_t^m (s_j) \right)^{a_{ijm}} \\
    &\quad \qquad  \cdot \left( \sum_{G'' \in \boldsymbol{\mathcal{G}}^m} P_{\boldsymbol{\pi}} \left( \mathbb{G}_{t}^{m} \left( {\boldsymbol \mu}_{t} \right) = G'' \mid x''_t = s_j \right)  \sum_{u \in \mathcal{U}} \pi_{t}^m \left(u \mid s_{j} \right) \cdot P \left( s_i \mid s_{j}, u, G'' \right) \right)^{a_{i j m}}
\end{align*}
which, for the power law running example, can be reformulated as
\begin{align*}
    &P_{\boldsymbol{\pi}, \boldsymbol{\mu}} \left( \mathbb{G}_{t + 1}^{k}  \left( {\boldsymbol \mu}_{t} \right) = G, x_{t+1} = x \right) \\
    &\qquad \approx \sum_{G' \in \boldsymbol{\mathcal{G}}^k} \sum_{x' \in \mathcal X} P_{\boldsymbol{\pi}, \boldsymbol{\mu}} \left( \mathbb{G}_{t}^{k}  \left( {\boldsymbol \mu}_{t} \right) = G', x_t = x' \right) 
    \left[ \sum_{u \in \mathcal{U}} \pi^k \left( u \mid x' \right) P \left( x \mid x', u, G'\right) \right] \\
    &\qquad \cdot \sum_{c \in \boldsymbol{\mathcal{C}}^k} \left[ \sum_{\boldsymbol{a}_2 \in \boldsymbol{\mathcal{A}}^k (G', c)} \prod_{j=1}^{d} \binom{g_j'}{a_{j 1}, \ldots, a_{j \infty}} \right. \\
    &\qquad \qquad \qquad \left. \cdot \prod_{m \in [k^*] \cup \{ \infty \}}  \left( \frac{m^{1 - \gamma} \mu_t^m (s_j)}{\left[ \sum_{\ell = k^* + 1}^{\infty} \ell^{1 - \gamma} \right] \mu_t^\infty (s_j) + \sum_{h=1}^{k^*} h^{1 - \gamma} \mu_t^h (s_j)} \right)^{a_{j m}} \right] \\
    &\cdot \sum_{\boldsymbol{a}_3 \in \boldsymbol{\mathcal{A}}^k_3 (G, G', c)} \frac{\prod_{j=1}^{d} \left( \mu_t^\infty (s_j)  \left( 1- \sum_{m' =1}^{k^*} \frac{(m')^{1- \gamma}}{\zeta (\gamma - 1)} \right) \right)^{a_{j \infty}}
    \prod_{m=1}^{k^*} \left( \frac{m^{1- \gamma} \mu_t^m (s_j)}{\zeta (\gamma - 1)}\right)^{a_{jm}}}{\sum_{\boldsymbol{a}_2' \in \boldsymbol{\mathcal{A}}^k_2 (G', c)} \prod_{j} \left( \mu_t^\infty (s_j)  \left( 1- \sum_{m' =1}^{k^*} \frac{(m')^{1- \gamma}}{\zeta (\gamma - 1)} \right) \right)^{a'_{j \infty}} \prod_{m=1}^{k^*} \left( \frac{m^{1- \gamma} \mu_t^m (s_j)}{\zeta (\gamma - 1)}\right)^{a'_{jm}}}\\
    &\qquad  \prod_{j=1}^{d} \prod_{m \in [k^*] \cup \{ \infty \}} \binom{\sum_i a_{i j m}}{a_{1 j m}, \ldots, a_{d j m}} \\
    &\quad \qquad \cdot \prod_{i=1}^{d} \left( \sum_{G'' \in \boldsymbol{\mathcal{G}}^m} P_{\boldsymbol{\pi}} \left( \mathbb{G}_{t}^{m} \left( {\boldsymbol \mu}_{t} \right) = G'' \mid x''_t = s_j \right)  \sum_{u \in \mathcal{U}} \pi_{t}^m \left(u \mid s_{j} \right) \cdot P \left( s_i \mid s_{j}, u, G'' \right) \right)^{a_{i j m}} \, .
\end{align*}
For notational convenience, define for each $j \in [d]$ and $m \in [k^*] \cup \{ \infty \}$
\begin{align*}
    p_{j m} \coloneqq \frac{P \left(\deg (v_1) = m \mid (v_0, v_1) \in E \right) \mu_t^m (s_j)}{P \left(\deg (v_1) > k^*\mid (v_0, v_1) \in E \right) \mu_t^\infty (s_j) + \sum_{k=1}^{k^*} P \left(\deg (v_1) = k \mid (v_0, v_1) \in E \right) \mu_t^k (s_j)}
\end{align*}
and for each $i,j \in [d]$ and $m \in [k^*] \cup \{ \infty \}$ 
\begin{align*}
    p_{i j m} &\coloneqq  P \left(\deg (v_1) = m \mid (v_0, v_1) \in E \right) \mu_t^m (s_j) \\
    &\qquad \qquad  \cdot  \sum_{G'' \in \boldsymbol{\mathcal{G}}^m} P_{\boldsymbol{\pi}} \left( \mathbb{G}_{t}^{m} \left( {\boldsymbol \mu}_{t} \right) = G'' \mid x''_t = s_j \right)  \sum_{u \in \mathcal{U}} \pi_{t}^m \left(u \mid s_{j} \right) \cdot P \left( s_i \mid s_{j}, u, G'' \right) \, .
\end{align*}
Then, the extensive approximation can be rewritten more compactly as
\begin{align*}
    &P_{\boldsymbol{\pi}, \boldsymbol{\mu}} \left( \mathbb{G}_{t + 1}^{k}  \left( {\boldsymbol \mu}_{t} \right) = G, x_{t+1} = x \right) \\
    &\qquad \approx \sum_{G' \in \boldsymbol{\mathcal{G}}^k} \sum_{x' \in \mathcal X} P_{\boldsymbol{\pi}, \boldsymbol{\mu}} \left( \mathbb{G}_{t}^{k}  \left( {\boldsymbol \mu}_{t} \right) = G', x_t = x' \right) 
    \left[ \sum_{u \in \mathcal{U}} \pi^k \left( u \mid x' \right) P \left( x \mid x', u, G'\right) \right] \\
    &\qquad \qquad \cdot \sum_{c \in \boldsymbol{\mathcal{C}}^k} \left[ \sum_{\boldsymbol{a}_2 \in \boldsymbol{\mathcal{A}}^k_2 (G', c)} \prod_{j=1}^{d} \binom{g_j'}{a_{j 1}, \ldots, a_{j \infty}} \prod_{m \in [k^*] \cup \{ \infty \}}  p_{j m}^{a_{j m}} \right] \\
    &\qquad \qquad \cdot \frac{\sum_{\boldsymbol{a}_3 \in \boldsymbol{\mathcal{A}}^k_3 (G, G', c)} \prod_{j=1}^{d} \prod_{m \in [k^*] \cup \{ \infty \}} \binom{\sum_i a_{i j m}}{a_{1 j m}, \ldots, a_{d j m}} \prod_{i=1}^{d} p_{i j m}^{a_{ijm}}}
    {\sum_{\boldsymbol{a}_2 \in \boldsymbol{\mathcal{A}}^k_2 (G', c)} \prod_{j, m} \left( P \left(\deg (v_1) = m \mid (v_0, v_1) \in E \right) \mu_t^m (s_j) \right)^{a_{j m}}} \, .
\end{align*}
Furthermore, we introduce
\begin{align*}
    \boldsymbol{p}_{2, j} \coloneqq \left( p_{j 1}, \ldots, p_{j k^*}, p_{j \infty} \right) \quad \mathrm{and} \quad \boldsymbol{a}_{2, j} \coloneqq \left( a_{j 1}, \ldots, a_{j k^*}, a_{j \infty} \right)
\end{align*}
for every $j \in [d]$ and similarly we define
\begin{align*}
    \boldsymbol{p}_{3, j m} \coloneqq \left( p_{1 j m}, \ldots, p_{d j m} \right) \quad \mathrm{and} \quad \boldsymbol{a}_{3, j m} \coloneqq \left( a_{1 j m}, \ldots, a_{d j m} \right)
\end{align*}
for every tuple $(j, m) \in [d] \times \left( [k^*] \cup \{ \infty \}\right)$. Then, the extensive approximation can be formulated as
\begin{align*}
    &P_{\boldsymbol{\pi}, \boldsymbol{\mu}} \left( \mathbb{G}_{t + 1}^{k}  \left( {\boldsymbol \mu}_{t} \right) = G, x_{t+1} = x \right) \\
    &\qquad \approx \sum_{G' \in \boldsymbol{\mathcal{G}}^k} \sum_{x' \in \mathcal X} \sum_{c \in \boldsymbol{\mathcal{C}}^k} P_{\boldsymbol{\pi}, \boldsymbol{\mu}} \left( \mathbb{G}_{t}^{k}  \left( {\boldsymbol \mu}_{t} \right) = G', x_t = x' \right) 
    \left[ \sum_{u \in \mathcal{U}} \pi^k \left( u \mid x' \right) P \left( x \mid x', u, G'\right) \right] \\
    &\qquad \quad \cdot \frac{ \left[ \sum_{\boldsymbol{a}_2 \in \boldsymbol{\mathcal{A}}^k_2 (G', c)} \prod_{j} \mathrm{Mult}_{\boldsymbol{p}_{2, j}} (\boldsymbol{a}_{2, j}) \right]
    \sum_{\boldsymbol{a}_3 \in \boldsymbol{\mathcal{A}}^k_3 (G, G', c)} \prod_{j, m} \mathrm{Mult}_{\boldsymbol{p}_{3, j m}} (\boldsymbol{a}_{3, j m})}
    {\sum_{\boldsymbol{a}_2 \in \boldsymbol{\mathcal{A}}^k_2 (G', c)} \prod_{j, m} \left( P \left(\deg (v_1) = m \mid (v_0, v_1) \in E \right) \mu_t^m (s_j) \right)^{a_{j m}}} \, .
\end{align*}

\section{Simulation Details} \label{sec:model_details}
We use MARLlib $1.0.0$ \citep{hu2023marllib} building on RLlib $1.8.0$ (Apache-2.0 license) \citep{liang2018rllib} and its PPO implementation \citep{schulman2017proximal} for IPPO and our algorithms. For our experiments, we used around $80 \, 000$ core hours on CPUs, and each training run usually took a single day of training on up to $96$ parallel CPU cores. For the policies we used two hidden layers of $256$ nodes with $\tanh$ activations. We used a discount factor of $\gamma=0.99$ with GAE $\lambda=1.0$, and training and minibatch sizes of $4000$ and $1000$, performing $5$ updates per training batch. The KL coefficient and clip parameter were set to $0.2$, with a KL target of $0.03$. The learning rate was set to $0.00005$. The problem details are found in the following.

\paragraph{Susceptible-Infected-Susceptible (SIS).}
In the SIS model with state space $\mathcal{X} \coloneqq \{ S, I \}$, agents are either infected ($I$) or susceptible to a virus ($S$). At each time step $t \in \mathcal{T}$, agents either protect themselves ($P$) or not ($\bar{P}$) which is formalized by the action space $\mathcal{U} \coloneqq \{ P, \bar{P} \}$. As usual, the game terminates at finite terminal time $T \in \mathbb{N}$ which can be interpreted as the time when a cure for the virus is found. Therefore, it remains to specify the transition dynamics. Susceptible agents who protect themselves at time $t$ also remain susceptible at time $t+1$, i.e.
\begin{align*}
    P^k (S \mid S, P, G)= 1 \quad \textrm{and} \quad  P^k (I \mid S, P, G)= 0,
\end{align*}
irrespective of their degree $k$ and neighborhood $G$. On the other hand, if a susceptible agent chooses action $\bar{P}$, the transition dynamics are
\begin{align*}
    P^k (I \mid S, \bar{P}, G) = \rho_{I} \cdot G(I) \cdot \left( \frac{2}{1 + \exp (-k/2)} -1 \right)
\end{align*}
and $P^k (S \mid S, \bar{P}, G) = 1 - P^k (I \mid S, \bar{P}, G)$, correspondingly, and where $\rho_{I} >0$ is a fixed infection rate. Apart from that, infected agents recover with some fixed recovery rate $1 \geq \rho_R \geq 0$, independent of their action and degree, which means that
\begin{align*}
    P^k (S \mid I, \bar{P}, G) = P^k (S \mid I, P, G) = \rho_{R} \, .
\end{align*}
To complete the model, the reward per agent taking action $u \in \mathcal{U}$ in state $x \in \mathcal{X}$ at each time $t$ is
\begin{align*}
    r(x,u) = - c_P \cdot \boldsymbol{1}_P (u) - c_I \cdot \boldsymbol{1}_I (x),
\end{align*}
where the cooperative objective $J$ is obtained by talking the average reward over all agents and summing up over all time points. Here, $c_P$ and $c_I$ denote the constant costs of protecting oneself and being infected, respectively.
In our experiments from the main text, the chosen parameter values are $\mu_0 (I) = 0.4, \mu_0 (S) = 0.6, T=50, \rho_I = 0.4, \rho_R = 0.1, c_P = 0.5$, and $c_I = 1$.

\paragraph{Susceptible-Infected-Recovered (SIR).}
In the SIR model, we extend the state space from the SIS by the recovered state $R$ and obtain $\mathcal{X} \coloneqq \{S, I, R \}$. As only infected agents can recover, the transition dynamics of the SIS model are modified by
\begin{align*}
     P^k (R \mid I, \bar{P}, G) = P^k (R \mid I, P, G) = \rho_{R}
\end{align*}
and 
\begin{align*}
     P^k (R \mid R, \bar{P}, G) = P^k (R \mid R, P, G) = 1,
\end{align*}
to formalize that recovered agents cannot become susceptible or infected again. The rewards and hence objective remain the same as in the SIS model. In the experiments, we set the parameter values $\mu_0 (I) = 0.1, \mu_0 (S) = 0.9, T=50, \rho_I = 0.1, \rho_R = 0.02, c_P = 0.25$, and $c_I = 1$.

\paragraph{Graph coloring (Color).}
In this problem, the state space consists of five colors $\mathcal{X} \coloneqq \{ x_1, x_2, x_3, x_4, x_5 \}$ allocated on a circle. Agents can move from the current color to the next color on the left ($\ell$), to the next one on the right ($r$), or stay at their current color ($s$) such that the action space is $\mathcal{U} \coloneqq \{ \ell, r, s \}$. The group of agents is also supposed to come close to a target distribution $\nu \in \mathcal{P} (\mathcal{X})$.
To keep notations manageable, we make the auxiliary definition
\begin{align*}
    \Tilde{G}_k \coloneqq \min(1, G^2 \cdot \rho_d \cdot \exp (-2/k)),
\end{align*}
where $\rho_d > 0$ is a constant noise factor.
The following three matrices specify the transition dynamics, where the row is the current agent color and the column is the next agent color:
\begin{align*}
    P^k (\cdot \mid \cdot, \ell, G) = 
    \begin{pmatrix}
\Tilde{G}_k (x_1)/2 & 0 & 0 & \Tilde{G}_k (x_1)/2 & 1 - \Tilde{G}_k (x_1) \\
1 - \Tilde{G}_k (x_2) & \Tilde{G}_k (x_2)/2 & 0 & 0 & \Tilde{G}_k (x_2)/2 \\
\Tilde{G}_k (x_3)/2 & 1 - \Tilde{G}_k (x_3) & \Tilde{G}_k (x_3)/2 & 0 & 0 \\
0 & \Tilde{G}_k (x_4)/2 & 1 - \Tilde{G}_k (x_4) & \Tilde{G}_k (x_4)/2 & 0 \\
0 & 0 & \Tilde{G}_k (x_5)/2 & 1 - \Tilde{G}_k (x_5) & \Tilde{G}_k (x_5)/2
\end{pmatrix}
\end{align*}
and
\begin{align*}
    P^k (\cdot \mid \cdot, s, G) = 
    \begin{pmatrix}
1 - \Tilde{G}_k (x_1) & \Tilde{G}_k (x_1)/2 & 0 & 0 & \Tilde{G}_k (x_1)/2 \\
\Tilde{G}_k (x_2)/2 & 1 - \Tilde{G}_k (x_2) & \Tilde{G}_k (x_2)/2 & 0 & 0 \\
0 & \Tilde{G}_k (x_3)/2 & 1 - \Tilde{G}_k (x_3) & \Tilde{G}_k (x_3)/2 & 0 \\
0 & 0 & \Tilde{G}_k (x_4)/2 & 1 - \Tilde{G}_k (x_4) & \Tilde{G}_k (x_4)/2  \\
\Tilde{G}_k (x_5)/2 & 0 & 0 & \Tilde{G}_k (x_5)/2 & 1 - \Tilde{G}_k (x_5)
\end{pmatrix}
\end{align*}
and 
\begin{align*}
    P^k (\cdot \mid \cdot, r, G) = 
    \begin{pmatrix}
\Tilde{G}_k (x_1)/2 & 1 - \Tilde{G}_k (x_1) & \Tilde{G}_k (x_1)/2 & 0 & 0  \\
0 & \Tilde{G}_k (x_2)/2 & 1 - \Tilde{G}_k (x_2) & \Tilde{G}_k (x_2)/2 & 0 \\
0 & 0 & \Tilde{G}_k (x_3)/2 & 1 - \Tilde{G}_k (x_3) & \Tilde{G}_k (x_3)/2 \\
\Tilde{G}_k (x_4)/2 & 0 & 0 & \Tilde{G}_k (x_4)/2 & 1 - \Tilde{G}_k (x_4) \\
1 - \Tilde{G}_k (x_5) & \Tilde{G}_k (x_5)/2 & 0 & 0 & \Tilde{G}_k (x_5)/2 
\end{pmatrix} \, .
\end{align*}
The reward in our graph coloring model is defined as
\begin{align*}
    r (x_j, u, G) \coloneqq  - \left( \boldsymbol{1}_{\ell} (u)+ \boldsymbol{1}_{r} (u) \right) \cdot c_m
    - (G (x_{j-1}) + G (x_{j+1})) \cdot c_d - \sum_{i =1}^5 \vert \mu (x_i) - \nu (x_i) \vert \cdot c_\nu,
\end{align*}
where $c_m, c_d, c_\nu > 0$ are the costs of moving, having neighbors with neighboring colors, and deviating from the target distribution $\nu$, respectively. In our experiments, we choose the parameters $\mu_0 = (1, 0, 0, 0, 0), \nu = (0.1, 0.2, 0.4, 0.2, 0.1), T=20, \rho_d = 0.9, c_m = 0.1, c_d = 0.5$, and $c_\nu = 1$.

\paragraph{Rumor.}
The state space $\mathcal{X} \coloneqq \{ I, A \}$ in the rumor model consists of the state $A$ where an agent is aware of a rumor and state $I$ where the agent does not know the rumor and is therefore ignorant of the rumor. Agents either spread the rumor $S$ or decide not to do so $\bar{S}$ which results in the action space $\mathcal{U} \coloneqq \{ S, \bar{S} \}$. Since the rumor spreading probability increases with the number of aware numbers who decide to spread the rumor, we work with the extended state space $\mathcal{X}' \coloneqq \mathcal{X} \cup \left( \mathcal{X} \times \mathcal{U} \right)$. Then, the transition dynamics are
\begin{align*}
    P^k ( (A, u) \mid A, u, G) = P^k ( A \mid (A, u), u, G) =1, \quad \forall u \in \mathcal{U}, G \in \boldsymbol{\mathcal{G}}^k, k \in \mathbb{N}
\end{align*}
meaning that aware agents remain aware, and furthermore
\begin{align*}
    P^k ( (I, u) \mid I, u, G) = 1, \quad \forall u \in \mathcal{U}, G \in \boldsymbol{\mathcal{G}}^k, k \in \mathbb{N}
\end{align*}
and
\begin{align*}
    P^k ( A \mid (I, u), u, G) &= \min \left( 1, \rho_A \cdot G ((A, S)) \cdot \left( \frac{2}{1 + \exp (-k/2)} -1 \right) \right) \\
    P^k ( I \mid (I, u), u, G) &= 1- P^k ( A \mid (I, u), u, G) \, .
\end{align*}
To complete the rumor model, the reward is given by
\begin{align*}
    r (x, u, G) = \boldsymbol{1}_{(A, S)} (x) \cdot \left( r_S \cdot G((I, S)) + r_S \cdot G((I, \bar S)) - c_S \cdot G((A, S)) - c_S \cdot G((A, \bar S)) \right) 
\end{align*}
for each agent, where we obtain the overall objective by averaging over the individual rewards. In our experiments, the parameters are chosen as $\mu_0 (A)= 0.1, \mu_0 (I) = 0.9, T=50, \rho_A = 0.3, c_S = 16$, and $r_S = 4$.

\end{document}